\documentclass[10pt, fleqn]{scrartcl}
\usepackage[british]{babel}
\usepackage{amsmath,amsfonts,amssymb,microtype,mathdefs,array,mathrsfs,amsthm,enum2,relsize,booktabs}
\usepackage[section]{thmdefs2}
\usepackage[hidelinks]{hyperref}

\setkomafont{pagehead}{\small\itshape}
\setkomafont{sectioning}{\bfseries}
\setkomafont{descriptionlabel}{\itshape}

\typearea{calc}

\makeatletter
\newcommand\m@thsm@ller[2]{\mbox{\relscale{0.91}$\m@th#1#2$}}
\let\smaller\undefined
\DeclareRobustCommand\smaller[1]{\relax\ifmmode{\mathpalette\m@thsm@ller{#1}}\else{\relscale{0.91}#1}\fi}
\makeatother

\DeclareRobustCommand*{\dom}{\qopname\relax o{dom}}
\DeclareRobustCommand*{\rng}{\qopname\relax o{rng}}

\newcommand*{\Set}{\mathsf{Set}}

\newcommand*{\sh}{\mathrm{sh}}

\newcommand*{\suc}{\mathrm{suc}}

\newcommand*{\pr}{\mathrm{pr}}

\newcommand*{\op}{\mathrm{op}}

\newcommand*{\cl}{\mathrm{cl}}
\newcommand*{\un}{\mathrm{un}}

\DeclareRobustCommand*{\step}{\qopname\relax o{step}}
\newcommand*{\ML}{\smaller{\mathrm{ML}}}
\newcommand*{\WMSO}{\smaller{\mathrm{WMSO}}}
\newcommand*{\MSO}{\smaller{\mathrm{MSO}}}
\newcommand*{\GSO}{\smaller{\mathrm{GSO}}}
\newcommand*{\CMSO}{\smaller{\mathrm{CMSO}}}
\newcommand*{\CGSO}{\smaller{\mathrm{CGSO}}}

\newcommand*{\CL}{\smaller{\mathrm{CL}}}
\newcommand*{\WCL}{\smaller{\mathrm{WCL}}}
\newcommand*{\FO}{\smaller{\mathrm{FO}}}

\newcommand*{\muaf}{\mu_{\mathrm{af}}}
\newcommand*{\mup}{\mu_\rmp}

\newcommand*{\one}{\textsf{1}}

\newcommand*{\emptyseq}{\smaller{\langle\rangle}}

\newcommand*{\?}{\kern .08em}

\newcommand\nmodels{\not\models}
\newcommand\medcircle{\bigcirc}
\newcommand\lsem{[\![}
\newcommand\rsem{]\!]}
\newcommand\rcirclearrowleft{\circlearrowleft}

\newcommand\upqed{\vskip-\baselineskip\vskip-\belowdisplayskip}

\begin{document}
\title{Automata for Enriched Trees and Applications}
\author{Achim Blumensath}
\maketitle

\begin{abstract}
We study trees where each successor set is equipped with some additional structure.
We introduce a family of automaton models for such trees and prove their equivalence to
certain fixed-point logics. As~a consequence we obtain characterisations of various
variants of monadic second-order logic in terms of automata and fixed-point logics.
Finally, we use our machinery to give a simplified proof of the Theorem of Muchnik
and we derive several variants of this theorem for other logics.
\end{abstract}


\section{Introduction}   

Translations between logical formulae and automata provide a versatile tool both for
applications (such as model-checking algorithms) and for purely theoretical considerations
(like, e.g., studying questions of expressive power).
To give a concrete example\?: two of the strongest decidability results in logic are proved
by automata-theoretic methods. The first one is the Theorem of Muchnik~\cite{Walukiewicz02}
which states that a certain operation preserves the decidability of monadic second-order
theories\?; the second one is a Theorem of Puppis~\cite{Puppis10} on the decidability of
certain trees with non-regular labelling.

In most settings, the logic corresponding to automata turns out to be
monadic second-order logic, while many weaker logics can be characterised by automata
of special types. But a closer look reveals that it would be more appropriate to
say that the logics corresponding to automata are certain fixed-point logics.
The fact that these logics have the same expressive power as monadic second-order logic
seems to be coincidental.

A systematic study of the correspondence between tree automata and fixed-point logics
can be found in the thesis of Carreiro~\cite{Carreiro15} building on previous work
of Walukiewicz~\cite{JaninWalukiewicz96,Walukiewicz02} and
Venema~\cite{Venema06,KupkeVenema08,KissigVenema09}.
In this article, we improve upon those results in two ways. First of all, the development
in~\cite{Carreiro15} is not uniform\?: each equivalence between a logic and an automaton model
is proved on a case-by-case basis using ad-hoc methods.
While all definitions and proofs share a similar structure, the details of each case
are different. We were able to abstract away these details and to derive a general proof
that allows us to derive all results of~\cite{Carreiro15} in a uniform way.

The second improvement concerns the classes of structures supported by our framework.
\cite{Carreiro15}~only considers trees (possibly infinite).
For some applications, this is not sufficient since they require working with trees
where the successors of each vertex are equipped with some additional structure.
This can simply be an left-to-right ordering of the successors,
or it can be more substantial like, e.g., a probability measure.
For the above mentioned Theorem of Muchnik, for instance, one works with trees
expanded by an arbitrary number of additional relations, with the restriction that
these relations only relate siblings.
Our translation generalises in a straightforward way to trees enriched with such an additional
structure. As~a consequence we are able to give characterisations of a variety of logics
via suitable automaton models that hold not only for trees, but also for such `enriched' trees.

Finally, let me mention that, while most of the constructions and arguments
used below can be considered standard, it was surprisingly tricky to get the details right.

The overview of this article is as follows.
We set up our framework in Sections \ref{Sect: transition systems}~and~\ref{Sect: logic}.
In the first of these sections, we introduce the class of enriched trees we are working with,
while the second one introduces a general notion of a logic and a way to extend such
logics by fixed-point operators.
Sections \ref{Sect: games},~\ref{Sect: automata}, and~\ref{Sect: projection} contain
our translation between logics and automata.
Section~\ref{Sect: games} recalls some results about parity games,
which are needed in Section~\ref{Sect: automata} to prove the correctness of our translation.
Section~\ref{Sect: projection} contains automata constructions corresponding to
various set quantifiers.
In Section~\ref{Sect: MSO}, we use the results from Section~\ref{Sect: automata}
to give automata-theoretic characterisations of many variants of monadic second-order logic.
Finally, Section~\ref{Sect: Muchnik} contains an application consisting of a new, much simplified
proof of the Theorem of Muchnik and of some variants of this theorem for other logics,
some of them already known, some new.

\section{Enriched Trees}   
\label{Sect: transition systems}

In order to model transition systems with arbitrary enrichments, it is convenient
to use a coalgebraic approach, similar to that one in~\cite{Venema06}.
That is, we choose a functor~$\bbS$ that returns
the set of all possible enriched successor structures and then we model a transition
system as a function $\suc : S \to \bbS S$ mapping each state $s \in S$ to the
structure of its successors.

Let us start by defining the class of functors we will use.
We will restrict ourselves to so called \emph{polynomial functors,} as these have particularly
nice properties and they cover all the applications we have in mind.
A polynomial functor maps a set~$X$ to a set of $X$-labelled objects.
The formal definition is as follows.
\begin{Def}
A functor $\bbS : \Set \to \Set$ is \emph{polynomial} if it is of the form
\begin{align*}
  \bbS X = \sum_{i \in I} X^{D_i},
  \quad\text{for fixed sets~$I$ and $D_i$, $i \in I$.}
\end{align*}
Such a functor maps a function $f : X \to Y$ to the function $\bbS f : \bbS X \to \bbS Y$
applying~$f$ to each label. Formally,
\begin{align*}
  \bbF f(s) := f \circ s\,, \quad\text{for } s : D_i \to X\,.
\end{align*}

Thus, elements of~$\bbS X$ are functions $s : D_i \to X$, for some~$i$.
We denote the domain of such a function by $\dom(s) := D_i$.

(b) Let $\bbS$~be a polynomial functor. Two elements $s,t \in \bbS X$ have
\emph{the same shape,} in symbols $s \simeq_\sh t$,
if $s,t$~correspond to the same index $i \in I$, i.e, if $s : D_i \to X$
and $t : D_i \to X$, for some $i \in I$.
\end{Def}
\begin{Exams}
(a) The functor $\bbS X := X^*$ returning the set of finite words over the alphabet~$X$
is polynomial since it can be written as
\begin{align*}
  \bbS X = \sum_{n<\omega} X^n.
\end{align*}

(b) Fixing a signature~$\Sigma$, the functor~$\bbS$ mapping a set~$X$ to the set of
all countable $X$-labelled $\Sigma$-structures is polynomial since
\begin{align*}
  \bbS X = \sum_\frakA X^A,
\end{align*}
where the sum ranges over all countable $\Sigma$-structures~$\frakA$
and $A$~denotes the universe of the structure~$\frakA$.

(c) Let $\calD$~be the set of all pairs $\langle A,\mu\rangle$ where $A$~is a finite set
and $\mu$~is a probability measure on~$A$.
There exists a polynomial functor
\begin{align*}
  \bbS X := \sum_{\langle A,\mu\rangle \in \calD} X^A
\end{align*}
mapping a set~$X$ to the set of all $X$-labellings of some $\langle A,\mu\rangle \in \calD$.
\end{Exams}

We can now define transition systems as $\bbS$-coalgebras for some polynomial functor~$\bbS$.
\begin{Def}
Let $\bbS : \Set \to \Set$ be a polynomial functor and $\Sigma \in \Set$ an alphabet.

(a)
A \emph{$\Sigma$-labelled $\bbS$-enriched transition system} is a structure of the form
\begin{align*}
  \frakS = \langle S,\suc,\lambda,v_0\rangle
\end{align*}
where $S$~is the set of \emph{states,} $v_0 \in S$ is the \emph{initial state,}
$\lambda : S \to \Sigma$ is a \emph{labelling} of the states, and $\suc : S \to \bbS S$ is
a function assigning a \emph{successor structure} $\suc(v) \in \bbS S$
to each state $v \in S$.
We call the elements of $\dom(\suc(v))$ \emph{directions} at the state~$v$.

As for polynomial functors, we will use the notation $\dom(s)$ to denote the set of states
of a transition system~$s$ and we represent~$s$ by the labelling function
$s : \dom(s) \to \Sigma$, leaving the successor function $\suc : \dom(s) \to \bbS\dom(s)$
implicit.

(b) A \emph{homomorphism} $\varphi : s \to t$ between transition systems $s$~and~$t$
is a function $\varphi : \dom(s) \to \dom(t)$ satisfying
\begin{align*}
  t(\varphi(v)) = s(v)
  \qtextq{and}
  \suc(\varphi(v)) = \bbS\varphi(\suc(v))\,,
  \quad\text{for all } v \in \dom(s)\,.
\end{align*}

(c) For a transition system~$\frakS$ and a state $v \in S$, we denote by~$\frakS,v$
the transition system obtained from~$\frakS$ by changing the initial state to~$v$.
\end{Def}

\begin{Exams}
(a) For ordinary `non-enriched' transition systems, we can use a functor of the form
\begin{align*}
  \bbS X := \sum_{\kappa < \lambda} X^\kappa,
\end{align*}
where $\lambda$~is some fixed cardinal and the sum ranges over all cardinals~$\kappa$
less than~$\lambda$.

(c) For Markov chains, we use a polynomial functor~$\bbS$ where the index set~$I$ is a
set of probability measures.

(d) In Section~\ref{Sect: Muchnik} below we will define an operation~$\frakA^*$ called
a Muchnik iteration. This operation constructs an infinite $\bbS$-enriched tree where
$\bbS$~returns a set of $\Sigma$-structures.
\end{Exams}

\begin{Rem}
(a)
We could model transition systems as coalgebras for the combined functor $\bbS \times \Sigma$.
But for our purposes it is more convenient to separate the successors and the labelling.

(b)
One shortcoming of the notion of a polynomial functor is the fact that the index set~$I$ --
which corresponds to the class of objects we want to label -- is a set and not a proper class.
Therefore we will frequently have to introduce arbitrary cut-offs for the class
of successor structures. For instance, instead of the class of all trees,
we can only use classes of finitely branching ones, or countably branching ones.
While inconvenient, this is usually not a problem since we can always choose the cut-off point
large enough to cover the transition system under consideration.
\end{Rem}

Trees can now be defined as unravellings of transition systems.
\begin{Def}
Let $s$~be an $\bbS$-enriched transition system.

(a) A (finite) \emph{path} in~$s$ is a finite sequence of the form
\begin{align*}
  v_0,d_0,v_1,d_1,\dots,v_{n-1},d_{n-1},v_n
\end{align*}
where $v_0,\dots,v_n \in \dom(s)$ are states, each $d_i \in \dom(\suc(v_i))$ is a direction
at~$v_i$, and $v_{i+1} = \suc(v_i)(d_i)$.

(b)
The \emph{unravelling} of~$s$ is the transition system
$\un(s)$ whose domain consists of all finite paths in~$s$ starting at the initial state of~$s$,
the labelling assigns to each path $\langle v_0,\dots,v_n\rangle$ the label $s(v_n)$
of the last state, and the successor function is defined by
$\suc(\langle v_0,\dots,v_n\rangle) \simeq_\sh \suc(v_n)$ and
\begin{align*}
  \suc(\langle v_0,\dots,v_n\rangle)(d) := \langle v_0,\dots,v_n,\,d,\,\suc(v_n)(d)\rangle\,.
\end{align*}

(c) $s$~is an \emph{$\bbS$-enriched tree} if it is isomorphic to its unravelling $\un(s)$.
We denote the \emph{root} of~$s$ by~$\emptyseq$ (the empty sequence),
and we write $\bbT_\bbS\Sigma$~for the set of all $\Sigma$-labelled $\bbS$-enriched trees.
\end{Def}

\section{Logics}   
\label{Sect: logic}

Since we are dealing with many different logics,
we will use an abstract notion of a logic introduced in~\cite{Blumensath21}.
\begin{Def}
(a) A \emph{logic} is a triple $\langle L,\calM,{\models}\rangle$
where $L$~is a set of \emph{formulae,} $\calM$~a set of~\emph{models,} and
${\models} \subseteq \calM \times L$ a \emph{satisfaction relation.}
Frequently, we denote a logic simply by its set~$L$ of formulae, leaving $\calM$~and~$\models$
implicit.

(b) A \emph{morphism of logics}
$\langle\lambda,\mu\rangle : \langle L,\calM,{\models}\rangle \to \langle L',\calM',{\models'}\rangle$
is a pair of functions $\lambda : L \to L'$ and $\mu : \calM' \to \calM$ satisfying
\begin{align*}
  M' \models \lambda(\varphi) \quad\iff\quad \mu(M') \models \varphi\,,
  \quad\text{for all } \varphi \in L \text{ and } M' \in \calM'\,.
\end{align*}
Usually, we denote both components of a morphism with the same identifier.
\end{Def}
\begin{Exam}
Let $\Sigma$~be a fixed signature and let $\calC_\Sigma$~be the class of
all countable $\Sigma$-structures. Then $\langle\FO[\Sigma],\calC_\Sigma,{\models}\rangle$
forms a logic where $\FO[\Sigma]$ denotes the set of all first-order formulae over the
signature~$\Sigma$.
\end{Exam}

As the preceding example shows, our notion of a logic requires us to fix a signature.
Frequently this is inconvenient since we would like to consider several different signatures
at once. Therefore we introduce families of logics parametrised by their signature.
\begin{Def}
(a) A \emph{family of logics} is a functor~$L$ from the category of finite sets to the
category of logics.

(b)
A~family of logics~$L$ is \emph{over} a functor~$\bbS$ if, for every set~$\Sigma$,
the class of models of~$L[\Sigma]$ is equal to $\bbS\Sigma$ and,
for every function $f : \Sigma \to \Gamma$, the morphism $L[f] : L[\Sigma] \to L[\Gamma]$
is of the form $L[f] = \langle\lambda,\mu\rangle$ with $\mu = \bbS f$.
\end{Def}
\begin{Exams}
For transition systems, we can use the following families of logics.

(a)
For each set~$Q$, we obtain a logic $\langle\MSO[Q],\calM_Q,{\models}\rangle$
where $\MSO[Q]$ is the set of monadic second-order formulae over the signature
$\{ E \} + \set{ P_q }{ q \in Q }$ (without free variables),
and $\calM_Q$~is the set of all transition systems of the form
$\frakS = \langle S,E,(P_q)_{q \in Q}\rangle$ where $E$~is the edge relation
and the~$P_q$ are unary predicates.

Given a function $f : Q \to Q'$, we obtain a morphism $\MSO[f] : \MSO[Q] \to \MSO[Q']$
mapping a formula $\varphi \in \MSO[Q]$ to the formula $\MSO[f](\varphi)$
obtained from~$\varphi$ by renaming every predicate~$P_q$ to~$P_{f(q)}$.
The corresponding function on transition systems maps a system
$\frakS = \langle S,E,(P_{q'})_{q' \in Q'}\rangle$ to the system
\begin{align*}
  \MSO[f](\frakS) := \langle S,E,(P_{f(q)})_{q \in Q}\rangle\,.
\end{align*}

(b) We obtain similar families $\FO[Q]$, $\WMSO[Q]$, $\mu\ML[Q]$,
for \emph{first-order logic, weak monadic second-order logic,} and
\emph{the modal $\mu$-calculus.}
(We will define these logics formally in Section~\ref{Sect: MSO} below.
Note that our notation $\mu\ML$ is slightly inconsistent with our notation
for the fixed-point logics $\mu L$ defined below.)
\end{Exams}

Below we will mostly work with families of logics over functors of the form
$\bbS \circ \PSet$ where $\bbS$~is a polynomial functor and $\PSet$ the (covariant)
power-set functor. This corresponds to models whose elements are labelled by sets of symbols.
In particular, logics whose models are $\Sigma$-structures are of this form since
each element in a structure can belong to several predicates.
\begin{Def}
Let $L$~be a family of logics over $\bbS \circ \PSet$ where $\bbS$~is some polynomial functor.

(a) A~formula $\varphi \in L[X]$ is \emph{monotone} in a symbol $x \in X$ if,
given two models $s \simeq_\sh s'$ satisfying
\begin{align*}
  s'(v) = s(v)
  \qtextq{or}
  s'(v) = s(v) \cup \{x\}\,,
  \quad\text{for all } v \in \dom(s)\,,
\end{align*}
we have
\begin{align*}
  s \models \varphi \qtextq{implies} s' \models \varphi\,.
\end{align*}

(a) A~formula $\varphi \in L[X]$ is \emph{antitone} in~$x$ if,
given two models $s \simeq_\sh s'$ satisfying
\begin{align*}
  s'(v) = s(v)
  \qtextq{or}
  s'(v) = s(v) \cup \{x\}\,,
  \quad\text{for all } v \in \dom(s)\,,
\end{align*}
we have
\begin{align*}
  s' \models \varphi \qtextq{implies} s \models \varphi\,.
\end{align*}
\upqed
\end{Def}
\begin{Exam}
A first-order formula~$\varphi$ over the signature~$\Sigma$
is monotone in a relation $R \in \Sigma$ if, and only if,
$\varphi$~is equivalent to some formula~$\varphi'$ where
every occurrence of~$R$ in~$\varphi'$ is under an even number of negation signs.
\end{Exam}

For some of the applications below, we will have to put additional restrictions on
certain symbols in a formula. For instance, we might require some of the symbols to
occur only positively in the formula and others only negatively.
For this reason, we add two more parameters to our logics, leading
to logics $L[X,U,V]$ parametrised by three sets\?: the set~$X$ of all symbols,
the set~$U$ of symbols with the additional restriction, and the set~$V$ of symbols with
the opposite/dual restriction.
\begin{Def}
Let $\calC$~be the category of all triples $\langle X,U,V\rangle$ of sets with $U,V \subseteq X$
where the morphisms $f : \langle X,U,V\rangle \to \langle X',U',V'\rangle$ are functions
$f : X \to X'$ satisfying
\begin{align*}
  x \notin U \Rightarrow f(x) \notin U'
  \qtextq{and}
  x \notin V \Rightarrow f(x) \notin V'\,.
\end{align*}
A family of logics \emph{with polarities} is a functor~$L$
from~$\calC$ to the category of logics.
We say that such a family is \emph{over} a functor~$\bbS$ if,
for every triple $\langle X,U,V\rangle$, the class of models of $L[X,U,V]$ is equal to~$\bbS X$
and, for every morphism $f : \langle X,U,V\rangle \to \langle X',U',V'\rangle$,
the morphism $L[f] : L[X] \to L[X']$ is of the form $L[f] = \langle\lambda,\mu\rangle$ with
$\mu = \bbS f$.
\end{Def}
\begin{Exam}
For first-order logic, we obtain a family of logics with polarities
where $\FO^+[X,U,V]$ contains all formulae $\varphi \in \FO[X]$ such that every predicate
in~$U$ occurs only positively in~$\varphi$ and every predicate in~$V$ only negatively.
\end{Exam}

Besides monotonicity, there are two other restrictions we are interested in.
\begin{Def}
Let $L$~be a family of logics over $\bbS \circ \PSet$.

(a) A formula $\varphi \in L[X]$ is \emph{jointly discrete} in a set of symbols $C \subseteq X$
if it is monotone in every $x \in C$ and if $s \models \varphi$ implies
$s' \models \varphi$, for some~$s'$ that is obtained from~$s$ by removing all but one occurrence
of the labels in~$C$. Formally, $s' \simeq_\sh s$ and that there are
$v_0 \in \dom(s)$ and $c \in C$ such that
\begin{align*}
  s'(v_0) = s(v_0) \setminus (C \setminus \{c\})
  \qtextq{and}
  s'(v) = s(v) \setminus C\,, \quad\text{for all } v \neq v_0\,.
\end{align*}

Similarly, $\varphi$ is \emph{jointly co-discrete} in~$C$
if it is monotone in every $x \in C$ and if, given a model~$s$,
\begin{align*}
  \forall v_0\forall c[s_{v_0,c} \models \varphi]
  \qtextq{implies}
  s \models \varphi\,,
\end{align*}
where $s_{v_0,c}$~is defined by $s_{v_0,c} \simeq_\sh s$ and
\begin{align*}
  s_{v_0,c}(v_0) = s(v_0) \cup (C \setminus \{c\})
  \qtextq{and}
  s_{v_0,c}(v) = s(v) \cup C\,, \quad\text{for all } v \neq v_0\,.
\end{align*}

By $L_\rmd[X,U,V]$ we denote the set of all $L[X]$-formulae that are
jointly discrete in~$U$ and jointly co-discrete in~$V$.

(b) A formula $\varphi \in L[X]$ is \emph{continuous} in a symbol $x \in X$ if it is
monotone in~$x$ and if $s \models \varphi$ implies $s' \models \varphi$, for some
$s'$~obtained from~$s$ by removing all but finitely many occurrences of the label~$x$.
Formally, $s' \simeq_\sh s$ and there is some finite set $P \subseteq \dom(s)$ such that
\begin{align*}
  s'(v) = \begin{cases}
            s(v)                 &\text{if } v \in P\,, \\
            s(v) \setminus \{x\} &\text{if } v \notin P\,.
          \end{cases}
\end{align*}

Similarly, $\varphi$ is \emph{co-continuous} in~$x$ if it is
monotone in~$x$ and if, given a model~$s$,
\begin{align*}
  \forall P[s_P \models \varphi]
  \qtextq{implies}
  s \models \varphi\,,
\end{align*}
where the quantifier ranges over all finite sets $P \subseteq \dom(s)$ and
the model~$s_P$ is defined by $s_P \simeq_\sh P$ and
\begin{align*}
  s_P(v) = \begin{cases}
             s(v)            &\text{if } v \in P\,, \\
             s(v) \cup \{x\} &\text{if } v \notin P\,.
           \end{cases}
\end{align*}

By $L_\rmc[X,U,V]$ we denote the set of all $L[X]$-formulae that are
continuous in every $x \in U$ and co-continuous in every $x \in V$.
\end{Def}
\begin{Exam}
(a) The $\FO$-formula
\begin{align*}
  \varphi := \forall xPx \land \exists yQy \land \exists yRy
\end{align*}
is jointly co-discrete in $\{P\}$, jointly discrete both in~$\{Q\}$ and in~$\{R$\},
but not jointly discrete in $\{Q,R\}$.

(b) The formula
\begin{align*}
  \psi := \exists^\infty xPx \land \exists yQy
\end{align*}
is continuous in~$Q$, but not in~$P$.
\end{Exam}

We can define the usual logical connectives in our abstract setting.
\begin{Def}
Let $L$~be a family of logics with polarities over $\bbS \circ \PSet$.

(a) A formula $\psi \in L[X,U,V]$ is a \emph{dual} of $\varphi \in L[X,V,U]$ if
\begin{align*}
  s^\op \models \psi \quad\iff\quad s \nmodels \varphi\,,
\end{align*}
where $s^\op(v) := X \setminus s(v)$.
In this case we write $\varphi^\op := \psi$.

(b) A formula $\vartheta \in L[X,U,V]$ is the \emph{conjunction} of two formulae
$\varphi,\psi \in L[X,U,V]$ if
\begin{align*}
  s \models \vartheta \quad\iff\quad s \models \varphi \text{ and } s \models \psi\,.
\end{align*}
In this case we write $\varphi \land \psi := \vartheta$.

(c) A formula $\vartheta \in L[X,U,V]$ is the \emph{disjunction} of two formulae
$\varphi,\psi \in L[X,U,V]$ if
\begin{align*}
  s \models \vartheta \quad\iff\quad s \models \varphi \text{ or } s \models \psi\,.
\end{align*}
In this case we write $\varphi \lor \psi := \vartheta$.
\end{Def}
\begin{Rem}
Note that the formulae $\varphi \lor \psi$, $\varphi \land \psi$, and $\varphi^\op$
are well-defined up to logical equivalence (if they exist).
Also note that the symbols in~$\varphi^\op$ have their polarity reversed.
\end{Rem}
\begin{Exams}
For the first-order logic, we have
\begin{align*}
  [Px \lor \exists y(Exy \land Qy)]^\op &= Px \land \forall y(Exy \lso Qy)\,, \\
  \bigl[\exists x[Px \land \forall y[y \neq x \lso Qy]\bigr]^\op
  &= \forall x[Px \lor \exists y[y \neq x \land Qy]]\,.
\end{align*}
\upqed
\end{Exams}

The following observations follow immediately from the definitions.
\begin{Lem}
Let $L$~be a family of logics for $\bbS \circ \PSet$.
\begin{enuma}
\item If $\varphi \in L$ is monotone in~$x$, so is~$\varphi^\op$.
\item $\varphi \in L$ is jointly discrete in~$C$ if, and only if, $\varphi^\op$~is
  jointly co-discrete in~$C$.
\item $\varphi \in L$ is continuous in~$x$ if, and only if, $\varphi^\op$~is co-continuous
  in~$x$.
\end{enuma}
\end{Lem}

Next, let us introduce a variant~$\mu L$ of the modal $\mu$-calculus where the modal operators
are defined by $L$-formulae.
\begin{Def}
Let $L$~be a family of logics with polarities over $\bbS \circ \PSet$.

(a) We denote by $\mu L$ the following variant of the modal $\mu$-calculus for
$\bbS$-enriched transition systems.
Given a set~$\Sigma$ of labels and two disjoint sets $X,Y$ of \emph{fixed-point variables,}
we denote by $\mu L[\Sigma;X,Y]$ the smallest set of formulae satisfying the following
conditions.
\begin{itemize}
\item $a \in \mu L[\Sigma;X,Y]$, for every $a \in \Sigma$.
\item $x \in \mu L[\Sigma;X,Y]$, for every $x \in X \cup Y$.
\item $\varphi,\psi \in \mu L[\Sigma;X,Y]$ implies
  $\varphi \land \psi, \varphi \lor \psi \in L[\Sigma;X,Y]$ and
  $\neg\varphi \in \mu L[\Sigma;Y,X]$.
\item Let $\Theta \subseteq \mu L[\Sigma;X,Y]$ be a finite set of formulae
  and let $\Theta_+$~be the set of all $\vartheta \in \Theta$ containing a symbol from~$X$ and
  $\Theta_-$~the set of all $\vartheta \in \Theta$ containing a symbol from~$Y$.
  For every $\varphi \in L[\Theta,\Theta_+,\Theta_-]$, we have
  ${\medcircle}\varphi \in \mu L[\Sigma;X,Y]$.
\item If $\varphi \in \mu L[\Sigma;X + \{x\},Y]$ is monotone in~$x$,
  then $\mu x.\varphi \in L[\Sigma;X,Y]$.
  Similarly, if $\varphi \in \mu L[\Sigma;X,Y + \{y\}]$ is monotone in~$y$,
  then $\nu y.\varphi \in \mu L[\Sigma;X,Y]$.
\end{itemize}
We will always tacitly assume that fixed-point variables used by distinct
fixed-point operators in a formula are distinct.

The semantics is defined as follows.
For a given $\bbS$-enriched transition system $\frakS = \langle S,\suc,\lambda\rangle$,
an $\mu L$-formula $\varphi(x_0,\dots,x_{n-1}) \in \mu L[\Sigma;X,Y]$, and values
$P_0,\dots,P_{n-1} \subseteq S$ for the free fixed-point variables
$x_0,\dots,x_{n-1} \in X \cup Y$, we define the set $\lsem\varphi\rsem_{\bar P} \subseteq S$
of states satisfying~$\varphi$ inductively as follows.
\begin{align*}
  \lsem a\rsem_{\bar P} &:= \lambda^{-1}(a)\,, \\
  \lsem x_i\rsem_{\bar P} &:= P_i\,, \\
  \lsem\varphi \land \psi\rsem_{\bar P} &:=
    \lsem\varphi\rsem_{\bar P} \cap \lsem\psi\rsem_{\bar P}\,, \\
  \lsem\varphi \lor \psi\rsem_{\bar P} &:=
    \lsem\varphi\rsem_{\bar P} \cup \lsem\psi\rsem_{\bar P}\,, \\
  \lsem\neg\varphi\rsem_{\bar P} &:= S \setminus \lsem\varphi\rsem_{\bar P}\,, \\
  \lsem{\medcircle}\psi\rsem_{\bar P} &:=
    \bigset{ v \in S }{ \bbS f(\suc(v)) \models \psi } \\
    &\qquad\text{where } f : S \to \PSet(\Theta) \text{ maps } v \in S \text{ to }
    \set{ \vartheta \in \Theta }{ v \in \lsem\vartheta\rsem_{\bar P} }\,, \\
  \lsem\mu x.\psi\rsem_{\bar P} &\text{ is the least fixed-point of the function} \\
    &\qquad F_\psi : \PSet(S) \to \PSet(S) : Q \mapsto \lsem\psi\rsem_{\bar PQ}\,, \\
  \lsem\nu x.\psi\rsem_{\bar P} &\text{ is the greatest fixed-point of } F_\psi\,.
\end{align*}
Finally, we set $\mu L[\Sigma] := \mu L[\Sigma;\emptyset,\emptyset]$ and
\begin{align*}
  \frakS,v \models \varphi \quad\defiff\quad v \in \lsem\varphi\rsem_\emptyseq\,.
\end{align*}

(b) A formula $\varphi \in \mu L$ is \emph{pure} if, for every subformula
of the form $\mu x.\psi$ or $\nu x.\psi$, we have
$\psi \in \mu L[\Sigma;X,\emptyset] \cup L[\Sigma;\emptyset,X]$, for some~$X$.
We denote the corresponding fragment of~$\mu L$ by $\mup L$.

(c) A formula $\varphi \in \mu L$ is \emph{alternation-free} if
\begin{itemize}
\item for every subformula of the form $\mu x.\psi$, we have
  $\psi \in \mu L[\Sigma;X,\emptyset]$, for some~$X$, and
\item for every subformula of the form $\nu x.\psi$, we have
  $\psi \in \mu L[\Sigma;\emptyset,X]$, for some~$X$.
\end{itemize}
We denote the corresponding fragment of~$\mu L$ by $\muaf L$.
\end{Def}
\begin{Exams}
Let $\rmE_1$~be the logic whose formulae are boolean combinations of statements
of the form
\begin{align*}
  \sfE C := \text{`There exists a position whose set of labels is equal to~$C$.'}\,,
\end{align*}
for $C \subseteq \Sigma$.
Similarly, let $\rmE_\omega$~be the logic consisting of boolean combinations of statements
of the form
\begin{align*}
  \sfE_k C := \text{`There exist at least~$k$ positions whose set of labels is equal to~$C$.'}\,,
\end{align*}
for $C \subseteq \Sigma$ and $k < \omega$.

(a)
The fixed-point extension~$\mu\rmE_1$ coincides with the standard modal $\mu$-calculus.
Using $L := \rmE_\omega$, we obtain a graded version~$\mu\rmE_\omega$ of the $\mu$-calculus.

(b) The $\muaf\rmE_1$-formula
\begin{align*}
  \varphi := \mu x\bigl[a \lor {\medcircle}[\sfE\{x\}]\bigr]
\end{align*}
checks whether there is a reachable vertex labelled by~$a$.

(c) The $\mup\rmE_1$-formula
\begin{align*}
  \varphi := \nu x.\mu y\bigl[\bigl(a \land {\medcircle}[\sfE\{x\}]\bigr)
                              \lor {\medcircle}[\sfE\{y\}]\bigr]
\end{align*}
checks for the existence of a path with infinitely many letters~$a$.

(c) The $\mup\rmE_\omega$-formula
\begin{align*}
  \varphi := \nu x.\mu y\bigl[{\medcircle}[\sfE_2\{x\}] \lor {\medcircle}[\sfE_1\{y\}]\bigr]
\end{align*}
checks for an embedding of the infinite binary tree.

(e) To state that a given tree has exactly one vertex labelled~$a$ we can use the
$\mu\FO_\rmc$-formula
\begin{align*}
  &\mu x.\bigl[\bigl(a \land {\medcircle}[\forall u.P_\vartheta(u)]\bigr)
              \land \bigl(\neg a \land {\medcircle}\bigl[
                      \exists u[P_x(u) \land
                                \forall v(v \neq u \lso P_\vartheta(v))]
                      \bigr]\bigr)\bigr]\,,
\end{align*}
with $\vartheta := \nu y.[\neg a \land {\medcircle}[\forall u.P_y(u)]]$.
\end{Exams}

Finally, let us introduce a normal form for $\mu L$-formulae that will come in handy below
in our translation of formulae into automata.
\begin{Def}
Let $\varphi \in \mu L[\Sigma;X,Y]$.

(a) $\varphi$~is in \emph{negation normal form} if the only negations in~$\varphi$
appear in a subformula of the form~$\neg a$ with $a \in \Sigma$.

(b) $\varphi$~is \emph{guarded} if, for each subformula of the form $\sigma x.\psi$ with
$\sigma \in \{\mu,\nu\}$, every occurrence of the variable~$x$ in~$\psi$ is inside
a subformula~${\medcircle}\vartheta$ starting with a modal operator.
\end{Def}

Let us start with an observation that is useful to convert a formula into negation normal form.
\begin{Rem}
(a) Fixed-point formulae satisfy the usual negation law.
\begin{align*}
  \neg\nu x.\varphi \equiv \mu x.\neg\varphi[x \mapsto \neg x]\,.
\end{align*}

(b) For modal operators, we obtain the relation
\begin{align*}
  \neg{\medcircle}\psi \equiv {\medcircle}\bbS\PSet c(\psi^\op)\,,
\end{align*}
where $c : \mu L \to \mu L$ is the function exchanging each label~$\vartheta$ by~$\neg\vartheta$.
For the proof, note that we have
\begin{align*}
  s \models \bbS c(\psi^\op)
  \quad\iff\quad
  \bbS c(s) \models \psi^\op
  \quad\iff\quad
  s \nmodels \psi\,.
\end{align*}
Hence,
\begin{align*}
  \lsem\neg{\medcircle}\psi\rsem_{\bar P}
  &= S \setminus \lsem{\medcircle}\psi\rsem_{\bar P} \\
  &= S \setminus \set{ v \in S }{ \bbS f(\suc(v)) \models \psi } \\
  &= \set{ v \in S }{ \bbS f(\suc(v)) \nmodels \psi } \\
  &= \set{ v \in S }{ \bbS f(\suc(v)) \models \bbS c(\psi^\op) }
   = \lsem{\medcircle}\bbS c(\psi^\op)\rsem_{\bar P}\,.
\end{align*}
\upqed
\end{Rem}

\begin{Lem}\label{Lem: normal form for mu L}
Let $L$~be a logic over $\bbS \circ \PSet$ that is closed under duals.
\begin{enuma}
\item $\mu L$~is closed under duals.
\item Every $\mu L$-formula is equivalent to one that is guarded and in negation normal form.
\end{enuma}
\end{Lem}
\begin{proof}
(a) Let $\varphi \in \mu L$. By~(b), we can assume that $\varphi$~is in negation normal form.
Then $\varphi^\op$~is the formula obtained from~$\varphi$ by
\begin{itemize}
\item replacing every disjunction by a conjunction and vice versa,
\item replacing every $\mu$-operator by a~$\nu$ and vice versa.
\end{itemize}

(b) To transform a given fromula~$\varphi$ into negation normal form we can use the two
laws from the above remark.
It therefore remains to show how to make a formula in negation normal form guarded.
For each subformula $\mu x.\psi$, let $\psi'$~be the formula obtained from~$\psi$ by replacing
every unguarded occurrence of~$x$ by $\mathsf{false}$. Then we have
\begin{align*}
  \lsem\psi'\rsem_{\bar PQ} \subseteq \lsem\psi\rsem_{\bar PQ}
                            \subseteq Q \cup \lsem\psi'\rsem_{\bar PQ}\,,
\end{align*}
which implies that
\begin{align*}
  \lsem\mu x.\psi'\rsem_{\bar P}
  \subseteq \lsem\mu x.\psi\rsem_{\bar P}
  \subseteq \lsem\mu x.(x \lor \psi')\rsem_{\bar P}
  = \lsem\mu x.\psi'\rsem_{\bar P}\,.
\end{align*}
Hence, $\mu x.\psi$ is equivalent to $\mu x.\psi'$.
For greatest fixed points $\nu x.\psi$, we can similarly replace all unguarded~$x$ by
$\mathrm{true}$ instead.
\end{proof}

\section{Parity Games}   
\label{Sect: games}

Let us recall some material about parity games that will be used below in the translation
of automata into formulae.
\begin{Def}
(a)
A \emph{parity game} is a games played by two players (Player~$\Diamond$ and Player~$\Box$)
who move a token along the edges of a directed graph.
The game starts in a fixed vertex of the graph and in each turn one of the players chooses
an outgoing edge along which to move the token. To determine the moving player,
we assign a player to each vertex of this graph. The player assigned
to the current vertex chooses the outgoing edge.

The game ends if one of the player cannot make a move (because there are no outgoing edges),
in which case that player looses. Otherwise, the choices of the players determine
an infinite path through the game, called a \emph{play} of the game.
To determine the winner of such an infinite play, we label each vertex by a number,
its \emph{priority,} and the least priority seen infinitely often during the play
determines the winner\?: Player~$\Diamond$ wins if it is even\?; otherwise Player~$\Box$ wins.

Formally, we represent a parity game as a structure of the form
$\calG = \langle V_\Diamond,V_\Box,E,\Omega\rangle$
where $V := V_\Diamond + V_\Box$ is the set of \emph{positions,}
$V_\Diamond$~are the positions for Player~$\Diamond$,
$V_\Box$~are the positions for Player~$\Box$,
$E \subseteq V \times V$ is the \emph{edge relation,} and $\Omega : V \to \omega$
the \emph{priority function.}
Given an infinite play $(v_n)_{n<\omega}$ of such a game, Player~$\Diamond$ wins if
$(v_n)_{n<\omega}$ satisfies the \emph{parity condition\?:}
\begin{align*}
  \liminf_{n<\omega} \Omega(v_n) \text{ is even.}
\end{align*}

(b) A \emph{(positional) strategy} for Player~$\tau$ in a parity game
($\tau \in \{\Diamond,\Box\}$) is a function $\sigma : V_\tau \to E$
assigning an outgoing edge to each position of Player~$\tau$.
Such a strategy is \emph{winning} if Player~$\tau$ wins every play where all of his
choices are according to~$\sigma$.

(c) A game~$\calG$ is \emph{positionally determined} if there exist
two positional strategies $\sigma_\Diamond$ and $\sigma_\Box$ and a partition
$V = W_\Diamond + W_\Box$ of the positions ($W_\Diamond$~and $W_\Box$ may be empty) such that
$\sigma_\Diamond$~is winning for Player~$\Diamond$, for every game that starts in some
position from~$W_\Diamond$, and similarly $\sigma_\Box$~is winning for Player~$\Box$,
for all starting positions in~$W_\Box$.
\end{Def}

The usefulness of parity games stems from the following well-known fact.
\begin{Thm}[Emerson, Jutla, Mostowski~\cite{EmersonJutla91,Mostowski91}]
Every parity game is positionally determined.
\end{Thm}

The second result we will need is the fact that we can compute the winning regions of a parity
game by a formula of the modal $\mu$-calculus $\mu\ML$.
To do so, we encode a parity game $\calG = \langle V_\Diamond,V_\Box,E,\Omega\rangle$
as a transition system with predicates $V_\Diamond,V_\Box,\Omega_k$, for each priority~$k$.
\begin{Def}
For a $\mu\ML$-formula~$\psi$ and $k < \omega$, we set
\begin{align*}
  \step \psi      &:= [V_\Diamond \land \Diamond\psi] \lor [V_\Box \land \Box\psi]\,, \\
  \mathrm{win}    &:= \sigma_0x_0\cdots\sigma_{k-1}x_{k-1}.
                        \step \Lor_{i<k}(\Omega_i \land x_i)\,,
\end{align*}
where $\sigma_k$~is equal to~$\mu$ if $k$~is odd, and equal to~$\nu$ if $k$~is even.
\end{Def}
\begin{Prop}[Emerson, Jutla~\cite{EmersonJutla91}]\label{Prop: defining the winning region}
For all $k < \omega$, the $\mu\ML$-formula~$\mathrm{win}$ defines the winning region
for Player~$\Diamond$ on all parity games with at most~$k$ priorities.
\end{Prop}

The games obtained from automata have a special form\?: the players strictly alternate in making
moves.
\begin{Def}
A parity game $\calG = \langle V_\Diamond,V_\Box,E,\Omega\rangle$ is \emph{strictly alternating}
if, for every edge $\langle u,v\rangle \in E$,
\begin{align*}
  u \in V_\Diamond \Leftrightarrow v \in V_\Box
  \qtextq{and}
  u \in V_\Diamond \Rightarrow \Omega(u) \leq \Omega(v)\,.
\end{align*}
\upqed
\end{Def}

For games of this special form, we obtain the following corollary to
Proposition~\ref{Prop: defining the winning region}.
\begin{Def}
For $k < \omega$, we set
\begin{align*}
  \mathrm{win}^2_k      &:= \sigma_0x_0\cdots\sigma_{k-1}x_{k-1}.
                              {\Diamond}{\Box} \Lor_{i<k}(\Omega_i \land x_i)\,,
\end{align*}
where $\sigma_k$~is equal to~$\mu$ if $k$~is odd, and equal to~$\nu$ if $k$~is even.
\end{Def}
\begin{Cor}\label{Cor: defining the winning region of a strictly alternating game}
For all $k < \omega$, the $\mu\ML$-formula~$\mathrm{win}^2_k$ defines the winning region
(restricted to positions of Player~$\Diamond$)
for Player~$\Diamond$ on all strictly alternating parity games with at most~$k$ priorities.
\end{Cor}

\section{Automata}   
\label{Sect: automata}

When defining the transition relation for an automaton working on trees whose branching degree
is unbounded, we cannot simply list all allowed states for the successors since this
would be an infinite amount of data. To obtain a finite automaton we need to
adopt some formalism that can be used to specify the transition relation in a finite way.
Following an idea of Walukiewicz~\cite{Walukiewicz02}, we use logical formulae for this task.
Generalising the work of Carreiro~\cite{Carreiro15}, we will show that we can translate
$\mu L$-formulae into automata where the transition relation is defined by $L$-formulae.
\begin{Def}\label{Def: automaton}
Let $\bbS : \Set \to \Set$ be a polynomial functor and $L$~a family of logics with polarities
over $\bbS \circ \PSet$.

(a) Given a formula $\varphi \in L[Q]$ we say that a state $q \in Q$ \emph{occurs}
in~$\varphi$ if $\varphi \notin L[Q \setminus \{q\}]$.

(b) Let $\delta : Q \times \Sigma \to L[Q]$ be a function.
We associate with~$\delta$ a directed graph whose set of vertices is~$Q$ and there is an
edge $p \to q$ if $q$~occurs in $\delta(p,a)$, for some $a \in \Sigma$.
We say that $q \in Q$ is \emph{reachable} from $p \in Q$ if this graph has a path from~$p$
to~$q$. A~\emph{component} of~$\delta$ is a strongly connected component of the associated graph.

(c)
An \emph{(alternating) $L$-automaton over\/ $\bbS$-enriched trees} is a tuple
\begin{align*}
  \calA = \langle Q,\Sigma,\delta,q_0,\Omega\rangle
\end{align*}
where $Q$~is a finite set of \emph{states,} $\Sigma$~is a finite \emph{input alphabet,}
$q_0 \in Q$ the \emph{initial state,} $\Omega : Q \to \omega$ a \emph{priority function,} and
$\delta : Q \times \Sigma \to L[Q]$ is the \emph{transition function,}
which we assume satisfies the following property\?:
for every state $q \in Q$, there exists two disjoint sets $C,D \subseteq Q$ such that
\begin{itemize}
\item $C \cup D$ is the component containing~$q$,
\item $\delta(q,a) \in L[Q,C,D]$, for all $a \in \Sigma$, and
\item $\delta(q,a)$ is monotone in all states $p \in C \cup D$.
\end{itemize}

(d)
A~\emph{run} of such an automaton~$\calA$ on an $\bbS$-enriched tree~$t$ is a function
$\rho : \dom(t) \to \PSet(Q \times Q)$ with the following properties.
(A~pair $\langle p,q\rangle \in \rho(v)$ represents the fact that (a copy of) the automaton
is in state~$p$ at the predecessor of~$v$ and in state~$q$ at the vertex~$v$ itself.)
Given a state $p \in Q$, we write
\begin{align*}
  \rho_{/p}(v) := \set{ q \in Q }{ \langle p,q\rangle \in \rho(v) }\,.
\end{align*}
We say that $\rho$~is a run if $\rho(\emptyseq) = \{\langle q_0,q_0\rangle\}$ and
\begin{align*}
  \bbS \rho_{/p}(\suc(v)) \models \delta(p,t(v))\,,
  \quad\text{for all }
    p \in \bigcup_{q \in Q} \rho_{/q}(v)\,.
\end{align*}

(e)
Let $\rho$~be a run and $\beta = (v_i)_{i<\omega}$ an infinite branch of~$t$.
A~\emph{trace} of~$\rho$ \emph{along}~$\beta$ is a sequence $(q_i)_{i<\omega}$ of states
starting with the initial state~$q_0$
such that
\begin{align*}
  \langle q_i,q_{i+1}\rangle \in \rho(v_{i+1})\,, \quad\text{for all } i < \omega\,.
\end{align*}

A run~$\rho$ is \emph{accepting} if all traces $(q_i)_i$ along every branch
satisfy the parity condition\?:
\begin{align*}
  \liminf_{i<\omega} \Omega(q_i) \text{ is even.}
\end{align*}
The \emph{language recognised} by~$\calA$ is the set
\begin{align*}
  \lsem\calA\rsem := \set{ t \in \bbT_\bbS\Sigma }{ \calA \text{ has an accepting run on } t }\,.
\end{align*}
\upqed
\end{Def}
\begin{Exams}
Let $\bbS X := X^*$ be the functor for finitely branching transition systems.

(a) The $\muaf\rmE_1$-formula
\begin{align*}
  \varphi := \mu x\bigl[a \lor {\medcircle}[\sfE_1 x]\bigr]
\end{align*}
checks whether there is a reachable vertex labelled by~$a$.
We can translate it into an $\rmE_1$-automaton with a single state~$q$ with priority
$\Omega(q) := 1$ where the transition function is
\begin{align*}
  \delta(q,a) := \mathsf{true}
  \qtextq{and}
  \delta(q,b) := \sfE_1 q\,.
\end{align*}

(b)
The following $\FO$-automaton recognises the language of all trees $t \in \bbT_\bbS\{a,b\}$
where every subtree contains at least one letter~$a$.
We use two states $p$~and~$q$, where $p$~looks for the letter~$a$ while $q$~checks the condition
for every subtree. The state~$q$ is initial, the transition function is
\begin{align*}
  \delta(p,a) &:= \mathsf{true}\,, \\
  \delta(p,b) &:= \exists xP_px\,, \\
  \delta(q,c) &:= \forall xP_qx \land \exists xP_px\,, \quad\text{for } c \in \{a,b\}\,,
\end{align*}
and the priorities are $\Omega(p) := 1$ and $\Omega(q) := 0$.

(c) The $\mup\rmE_1$-formula
\begin{align*}
  \varphi := \nu x.\mu y\bigl[\bigl(a \land {\medcircle}[\sfE_1 x]\bigr)
                              \lor {\medcircle}[\sfE_1 y]\bigr]
\end{align*}
checks for the existence of a path with infinitely many letters~$a$.
We can translate it into an $\rmE_1$-automaton with two states $q_a$~and~$q_b$.
The priorities are $\Omega(q_a) := 0$ and $\Omega(q_b) := 1$ and the transition function is
\begin{align*}
  \delta(q_c,d) := \sfE_1 q_d\,, \quad\text{for } c,d \in \{a,b\}\,.
\end{align*}
\upqed
\end{Exams}

It will turn out that the two fragments $\mup L$ and $\muaf L$ of $\mu L$ can be characterised
by $L$-automata of the following form.
\begin{Def}
Let $\calA = \langle Q,\Sigma,\delta,q_0,\Omega\rangle$ be an $L$-automaton.

(a) $\calA$~is \emph{pure} if, for every component $C \subseteq Q$ of~$\calA$, we have
\begin{alignat*}{-1}
  \delta(q,a) &\in L[Q,C,\emptyset]\,, &&\quad\text{for all } q \in C \text{ and } a \in \Sigma\,, \\
\prefixtext{or}
  \delta(q,a) &\in L[Q,\emptyset,C]\,, &&\quad\text{for all } q \in C \text{ and } a \in \Sigma\,.
\end{alignat*}

(b) $\calA$~is \emph{weak} if
\begin{align*}
  \Omega(p) \leq \Omega(q)\,, \quad\text{for all states } q \text{ that occur in }
    \delta(p,a) \text{ for some } a \in \Sigma\,,
\end{align*}
and
\begin{align*}
  \Omega(q) \text{ is odd}  &\quad\Rightarrow\quad \delta(q,a) \in L[Q,C,\emptyset]\,, \\
  \Omega(q) \text{ is even} &\quad\Rightarrow\quad \delta(q,a) \in L[Q,\emptyset,C]\,,
\end{align*}
where $C \subseteq Q$ is the component of~$\calA$ containing~$q$.
\end{Def}
\begin{Rem}
Every weak $L$-automaton is equivalent to a (non-weak) $L$-automaton that
only uses the priorities $0$~and~$1$. (We can just replace the priority function~$\Omega$ by
the function $\Omega'(q) := \Omega(q) \bmod 2$.)
\end{Rem}

The main result of this section is the following equivalence between automata and
formulae. In fact, the two formalisms are close enough that, similar to the
modal $\mu$-calculus, we can consider automata a normal form for formulae.
\begin{Thm}\label{Thm: equivalent automata and fixed-point formulae}
Let $L$~be a family of logics with polarities over\/~$\bbS$ that is closed under
finite disjunctions, finite conjunctions, and duals.
\begin{enuma}
\item A~language $K \subseteq \bbT_\bbS\Sigma$ is $\mu L$-definable if, and only if,
  it is recognised by an $L$-automaton.
\item A~language $K \subseteq \bbT_\bbS\Sigma$ is $\mup L$-definable if, and only if,
  it is recognised by a pure $L$-automaton.
\item A~language $K \subseteq \bbT_\bbS\Sigma$ is $\muaf L$-definable if, and only if,
  it is recognised by a weak $L$-automaton.
\end{enuma}
Furthermore, the above translations between formulae and automata are effective
provided that disjunctions, conjunctions, and duals of $L$-formulae are computable.
\end{Thm}
The remainder of this section is devoted to the proof, which is basically the same
as the corresponding proof for the modal $\mu$-calculus. We only have to additionally check
that the transition functions and modal operators we construct are well-formed and
of the correct type. We split the proof into
Propositions \ref{Prop: automaton -> formula}~and~\ref{Prop: formula -> automaton} below.

Before doing so, it is useful to give an alternative definition of acceptance via a parity game.
\begin{Def}
Let $\calA = \langle Q,\Sigma,\delta,q_0,\Omega\rangle$ be an $L$-automaton
and $t \in \bbT_\bbS\Sigma$ an input tree. The \emph{acceptance game}
for~$\calA$ on~$t$ is the parity game
$\calG(\calA,t) := \langle V_\Diamond,V_\Box,E,\Omega'\rangle$ with positions
\begin{align*}
  V_\Diamond &:= \dom(t) \times Q\,, \\
  V_\Box     &:= \set{ \langle v,s\rangle \in \dom(t) \times \bbS\PSet(Q) }
                     { s \simeq_\sh \suc(v) }\,.
\end{align*}
The priorities are
\begin{align*}
  \Omega'(\langle v,q\rangle) := \Omega(q)
  \qtextq{and}
  \Omega'(\langle v,s\rangle) := \max \rng \Omega\,,
\end{align*}
for $v \in \dom(t)$, $q \in Q$, and $s \in \bbS Q$.

Finally, the edge relation is defined as follows.
Let $u,v \in \dom(t)$, $q \in Q$, and $s \in \bbS\PSet(Q)$.
There are edges
\begin{alignat*}{-1}
  \langle v,q\rangle &\to \langle v,s\rangle
  &&\quad\defiff\quad
  &&s \models \delta(q,t(v))\,, \\[0.5em]
  \langle v,s\rangle &\to \langle u,q\rangle
  &&\quad\defiff\quad
  &&u = \suc(v)(d) \qtextq{and} q \in s(d)\,, \\
  &&&&&\quad\text{for some } d \in \dom(\suc(v))\,.
\end{alignat*}
\upqed
\end{Def}

\begin{Prop}\label{Prop: game characterisation of acceptance}
An $L$-automaton~$\calA$ accepts a tree~$t$ if, and only if,
Player~$\Diamond$ has a winning strategy in the game $\calG(\calA,t)$.
\end{Prop}
The proof is entirely standard\?: every accepting run can be used to define a winning strategy
and every winning strategy an accepting run.

For the two directions of the proof of
Theorem~\ref{Thm: equivalent automata and fixed-point formulae},
we generalise the standard translation between the modal $\mu$-calculus and tree automata.
We start with the translation of automata into formulae. To simplify the construction
we will use a variant of~$\mu L$ with \emph{simultaneous} fixed points.
\begin{Def}
The variant of $\mu L$ with \emph{simultaneous fixed points} has fixed-point formulae
of the form
\begin{align*}
  \mu_k\bar x.\bar\psi
  \qtextq{and}
  \nu_k\bar x.\bar\psi\,,
\end{align*}
where $\bar x$~is an $n$-tuple of (pairwise distinct) fixed-point variables,
$\bar\psi$~an $n$-tuple of $\mu L$-formulae that are monotone in the variables~$\bar x$,
and $k < n$ is an index.
The semantics $\lsem\mu_k\bar x.\bar\psi\rsem_{\bar P}$ of such a formula is defined
as follows. Let $\bar T$~be the least fixed point of the operation
$F : \PSet(S)^n \to \PSet(S)^n$ defined by
\begin{align*}
  F(\bar Q) := \bigl\langle\lsem\psi_k\rsem_{\bar P\bar Q}\bigr\rangle_{k < n}\,.
\end{align*}
Then $\lsem\mu_k\bar x.\bar\psi\rsem_{\bar P} := T_k$.
\end{Def}
\begin{Lem}\label{Lem: elimination of simultaneous fixed points}
Every $\mu L[\Sigma;X,Y]$-formula with simultaneous fixed points can be translated to one
without.
Furthermore, if the given formula is alternation-free or pure, so is the resulting formula.
\end{Lem}
\begin{proof}
This is a standard construction, which we will recall for convenience.
Consider a formula of the form $\varphi = \mu_k\bar x.\bar\psi$ where $\bar x$~and~$\bar\psi$
are $n$-tuples. (The case of a greatest fixed point is handled analogously.)
We may assume by induction that the formulae~$\psi_i$ do not contain simultaneous fixed points.
We transform~$\varphi$ into a $\mu L$-formula in two steps.

First, we modify the formula such the variables $x_0,\dots,x_{i-1}$ are not free in~$\psi_i$,
for $i<n$. To do so, we replace each occurrence of~$x_0$ in~$\psi_i$, for $i > 0$, by
the formula $\mu x_0.\psi_0$.
Next we replace each~$x_1$ in (the modified version of)~$\psi_i$, for $i > 1$,
by $\mu x_1.\psi_1$
Continuing this way, we replace each~$x_j$ in~$\psi_i$, for $i > j$, by $\mu x_j.\psi_j$.

We denote the resulting formulae again by $\psi_0,\dots,\psi_{n-1}$.
In the second step, we eliminate the remaining variables. We start with the formula
\begin{align*}
  \varphi_{n-1} := \mu x_{n-1}.\psi_{n-1}\,.
\end{align*}
Next, we construct
\begin{align*}
  \varphi_{n-2} := \mu x_{n-2}.\psi'_{n-2}\,,
\end{align*}
where $\psi'_{n-2}$ is the formula obtained from~$\psi_{n-2}$ by replacing each
occurrence of~$x_{n-1}$ by $\varphi_{n-1}$.
Continuing in this way, we set
\begin{align*}
  \varphi_j := \mu x_j.\psi'_j\,,
\end{align*}
where $\psi'_j$~is the formula obtained from~$\psi_j$ by replacing each
occurrence of~$x_i$ with $i > j$ by~$\varphi_i$.
The resulting formulae $\varphi_0,\dots,\varphi_{n-1}$ belong to $\mu L$
and define the respective components of the fixed point.
In particular, $\varphi_k$~is equivalent to $\mu_k\bar x.\bar\psi$.

Finally, note that our construction preserves purity and alternation freeness.
\end{proof}

The two directions of the proof of
Theorem~\ref{Thm: equivalent automata and fixed-point formulae}
can now be proved as follows.
\begin{Prop}\label{Prop: automaton -> formula}\leavevmode
\begin{enuma}
\item Every language recognised by an $L$-automaton is $\mu L$-definable.
\item Every language recognised by a weak $L$-automaton is $\muaf L$-definable.
\item Every language recognised by a pure $L$-automaton is $\mup L$-definable.
\end{enuma}
\end{Prop}
\begin{proof}
Let $\calA = \langle Q,\Sigma,\delta,q_0,\Omega\rangle$ be an $L$-automaton,
$t \in \bbT_\bbS\Sigma$ an input tree, and $\calG = \langle V_\Diamond,V_\Box,E,\Omega\rangle$
the associated acceptance game.
It is sufficient to find a formula~$\varphi$ of the respective logic such that
\begin{align*}
  t \models \varphi \quad\iff\quad
  \text{the initial position of } \calG \text{ is winning for Player~$\Diamond$.}
\end{align*}

More precisely, for each state $q \in Q$, we will construct a formula $\varphi_q \in \mu L$
such that, for every vertex $v \in \dom(t)$,
\begin{align*}
  t,v \models \varphi_q
  \quad\iff\quad
  \text{Player~$\Diamond$ wins when starting in the position } \langle v,q\rangle\,.
\end{align*}
We proceed by induction on the number of components of~$\calA$ that are reachable from~$q$.
Hence, let $C$~be a component with $q \in C$ and let $P \subseteq Q$
be the set of all states $p \in Q \setminus C$ reachable from some $q \in C$.
By inductive hypothesis, we already know the formulae~$\varphi_p$, for $p \in P$.
By Corollary~\ref{Cor: defining the winning region of a strictly alternating game},
the $\mu\ML$-formula
\begin{align*}
  \mathrm{win}^2 := \sigma_0x_0\cdots\sigma_{k-1}x_{k-1}.
                      {\Diamond}{\Box} \Lor_{i<k}(\Omega_i \land x_i)
\end{align*}
defines the winning region for Player~$\Diamond$ in~$\calG$.
It follows that the $\mu\ML$-formula
\begin{align*}
  \chi_C := \sigma_0x_0\cdots\sigma_{k-1}x_{k-1}.
               {\Diamond}{\Box}
                 \Bigl[\Lor_{p \in C}(Q_p \land x_{\Omega(p)}) \lor
                         \Lor_{p \in P} (Q_p \land \chi_P)\Bigr]
\end{align*}
defines the winning region restricted to positions with a state in~$C$.
($\chi_P$~is the corresponding formula for states in~$P$ and $Q_p$~is the predicate checking
that the current state is~$p$. Hence, $\Omega_k \equiv \Lor_{p \in \Omega^{-1}(k)} Q_p$.)

(a),~(b)
We will inductively translate every subformula~$\psi(x_0,\dots,x_{n-1})$ of~$\chi_C$ into an
$\mu L$-formula $\psi^*_q(\bar x_0,\dots,\bar x_{n-1})$ such that, for all $v \in \dom(t)$ and
$P_0,\dots,P_{k-1} \subseteq V_\Diamond$,
\begin{align*}
  \calG,\langle v,q\rangle \models \psi(P_0,\dots,P_{k-1})
  \quad\iff\quad
  t,v \models \psi^*_q(\bar P_0,\dots,\bar P_{k-1})\,,
\end{align*}
where $\bar P_i := (P_i^p)_{p \in Q}$ with
\begin{align*}
  P_i^p := \set{ v \in \dom(t) }{ \langle v,p\rangle \in P_i }\,.
\end{align*}
Note that to each variable~$x_i$ in~$\chi_C$ there corresponds
a $Q$-tuple $(x_{i,q})_{q \in Q}$ in the translation.
We start with
\begin{align*}
  \Bigl[\Lor_{p \in C}(Q_p \land x_{\Omega(p)}) \lor
          \Lor_{p \in P} (Q_p \land \chi_P)\Bigr]^*_q
    := \begin{cases}
         x_{\Omega(q),q} &\text{if } q \in C\,, \\
         \varphi_q       &\text{if } q \in P\,.
       \end{cases}
\end{align*}
For the modal operators, we set
\begin{align*}
  \Bigl[{\Diamond}{\Box}\Bigl(\Lor_{p \in C}(Q_p \land x_{\Omega(p)}) \lor
          \Lor_{p \in P} (Q_p \land \chi_P)\Bigr)\Bigr]^*_q
    := \Lor_{a \in \Sigma}
         \bigl[a \land {\medcircle}\bbS f(\delta(q,a))\bigr]\,,
\end{align*}
where $X := \set{ x_{\Omega(p),p} }{ p \in C }$
and the function $f : Q \to \mu L[\Sigma;X,X]$ is defined by
\begin{align*}
  f(p) := \begin{cases}
            x_{\Omega(p),p} &\text{if } p \in C\,, \\
            \varphi_p       &\text{if } p \in P\,.
          \end{cases}
\end{align*}
Note that the above formula is well-formed since we have $\delta(q,a) \in L[Q,C_1,C_2]$,
for some partition $C = C_1 + C_2$ independent of~$a$, which implies that
\begin{align*}
  {\medcircle}\bbS f(\delta(q,a)) \in \mu L[\Sigma;X_1,X_2]
  \qtextq{where}
  X_i := \set{ x_{\Omega(p),p} }{ p \in C_i }\,.
\end{align*}

Furthermore, in case~(b), we have
$\delta(q,a) \in L[Q,C,\emptyset] \cup L[Q,\emptyset,C]$, which implies that
\begin{align*}
  {\medcircle}\bbS f(\delta(q,a)) \in
    \mup L[\Sigma;X,\emptyset] \cup \mup L[\Sigma;\emptyset,X]\,.
\end{align*}
Hence, the resulting formula is pure.

Finally, we translate the fixed-point operators using simultaneous fixed-points as
\begin{alignat*}{-1}
  (\mu x_i.\psi)^*_q &:=
    \mu_q (x_{i,p})_{p \in C \cap \Omega^{-1}(i)}.(\psi^*_p)_{p \in C \cap \Omega^{-1}(i)}\,,
    &&\quad\text{if $i$ is odd,} \\
\prefixtext{or}
  (\nu x_i.\psi)^*_q &:=
    \nu_q (x_{i,p})_{p \in C \cap \Omega^{-1}(i)}.(\psi^*_p)_{p \in C \cap \Omega^{-1}(i)}\,,
    &&\quad\text{if $i$ is odd.}
\end{alignat*}
Note that these formulae are well-formed since every state $p \in C$ occurs positively
in~$\delta(q,a)$. This implies that every variable~$x_{i,p}$ occurs positively in~$\psi_{p'}$,
for $p,p' \in C \cap \Omega^{-1}(i)$.

It remains to prove the correctness of our translation. We proceed by induction on the
formula~$\psi$. Since most steps are straightforward, we only consider the case of the
modal operator. Hence, suppose that $\psi = {\Diamond}{\Box}\vartheta$.
By definition of~$\calG$, we have
\begin{align*}
  \calG,\langle v,q\rangle \models {\Diamond}{\Box}\vartheta
\end{align*}
if, and only if, there exists a labelling $s \in \bbS Q$ such that $s \simeq_\sh \suc(v)$,
\begin{align*}
  s \models \delta(q,t(v))
  \qtextq{and}
  \calG,\langle\suc(v)(d),s(d)\rangle \models \vartheta\,, \quad\text{for all } d \in \dom(\suc(v))\,.
\end{align*}
By inductive hypothesis, this is equivalent to the existance of some $s \in \bbS Q$ satisfying
\begin{align*}
  s \models \delta(q,t(v))
  \qtextq{and}
  t,\suc(v)(d) \models \vartheta^*_{s(d)}\,, \quad\text{for all } d \in \dom(\suc(v))\,.
\end{align*}
This last statement holds if, and only if,
\begin{align*}
  t,v \models
    {\medcircle}\bbS f(\delta(q,t(v)))_{p \in Q}(\bar P_0,\dots,\bar P_{n-1})\,,
\end{align*}
which is equivalent to
\begin{align*}
  t,v \models
    \Lor_{a \in \Sigma}
      \bigl[a \land {\medcircle}\bbS f(\delta(q,a))_{p \in Q}(\bar P_0,\dots,\bar P_{n-1})\bigr]\,.
\end{align*}

(c) As above, we can use the $\mu\ML$-formula
\begin{align*}
  \chi_C := \sigma_0x_0\cdots\sigma_{k-1}x_{k-1}.
               {\Diamond}{\Box}
                 \Bigl[\Lor_{p \in C}(Q_p \land x_{\Omega(p)}) \lor
                         \Lor_{p \in P} (Q_p \land \chi_P)\Bigr]
\end{align*}
to define the winning region restricted to positions with a state in~$C$.
Since the automaton is weak, all states in~$C$ have the same priority~$k$.
Consequently, the formula simplifies to
\begin{align*}
  \chi'_C := \sigma_kx.
               {\Diamond}{\Box}
                 \Bigl[\Lor_{p \in C}(Q_p \land x) \lor
                         \Lor_{p \in P} (Q_p \land \chi'_P)\Bigr]\,.
\end{align*}
As above, we inductively translate every subformula~$\psi$ of~$\chi'_C$ into
into an $\mup L$-formula~$\psi^*_q$, starting with
\begin{align*}
  \Bigl[\Lor_{p \in C}(Q_p \land x) \lor \Lor_{p \in P} (Q_p \land \chi'_P)\Bigr]^*_q
    := \begin{cases}
         x_q       &\text{if } q \in C\,, \\
         \varphi_q &\text{if } q \in P\,,
       \end{cases}
\end{align*}
and
\begin{align*}
  \Bigl[{\Diamond}{\Box}\Bigl(\Lor_{p \in C}(Q_p \land x_{\Omega(p)}) \lor
          \Lor_{p \in P} (Q_p \land \chi'_P)\Bigr)\Bigr]^*_q
    := \Lor_{a \in \Sigma}
         \bigl[a \land {\medcircle}\bbS f(\delta(q,a))\bigr]\,,
\end{align*}
where $X := \set{ x_p }{ p \in C }$ and the function $f : Q \to \mup L[\Sigma;X,X]$ is defined by
\begin{align*}
  f(p) := \begin{cases}
            x_p       &\text{if } p \in C\,, \\
            \varphi_p &\text{if } p \in P\,.
          \end{cases}
\end{align*}
Note that this formula is well-formed since, depending on whether or not $k$~is odd,
we have $\delta(q,a) \in L[Q,C,\emptyset]$ or $\delta(q,a) \in L[Q,\emptyset,C]$,
which implies that
\begin{align*}
  {\medcircle}\bbS f(\delta(q,a)) \in
    \begin{cases}
      \mup L[\Sigma;X,\emptyset] &\text{if $k$ is odd}\,, \\
      \mup L[\Sigma;\emptyset,X] &\text{if $k$ is odd}\,.
    \end{cases}
\end{align*}
Finally, we translate the fixed-point operator by
\begin{alignat*}{-1}
  (\mu x.\psi)^*_q &:= \mu_q (x_p)_{p \in Q}.(\psi^*_p)_{p \in Q}\,,
    &&\quad\text{if $k$ is odd,} \\
\prefixtext{or}
  (\nu x.\psi)^*_q &:= \nu_q (x_p)_{p \in Q}.(\psi^*_p)_{p \in Q}\,,
    &&\quad\text{if $k$ is odd,}
\end{alignat*}
which is alternation-free.
\end{proof}

\begin{Exam}
To understand the construction in the following proof, let us consider the $\mu\rmE_1$-formula
\begin{align*}
  \varphi := \nu x.\mu y\bigl[\bigl(a \land {\medcircle}[\sfE x]\bigr)
                              \lor {\medcircle}[\sfE y]\bigr]
\end{align*}
which checks for the existence of a path with infinitely many letters~$a$.
This formula has the following subformulae.
\begin{alignat*}{-1}
  &a\,,\qquad &&x\,,\qquad
  &\psi_0 &:= {\medcircle}[\sfE y]\,, \qquad
  &\psi_2 &:= a \land \psi_1\,, \qquad
  &\psi_4 &:= \mu y.\psi_3\,, \\
  &&&y\,,\qquad
  &\psi_1 &:= {\medcircle}[\sfE x]\,, \qquad
  &\psi_3 &:= \psi_2 \land \psi_0\,, \qquad
  &\varphi\,.
\end{alignat*}
We translate~$\varphi$ into the automaton with states
\begin{align*}
  Q := \{x, y, \varphi\}\,,
\end{align*}
priorities
\begin{align*}
  \Omega(\varphi) := 0\,, \quad
  \Omega(x) := 0\,, \quad
  \Omega(y) := 1\,,
\end{align*}
and transitions
\begin{align*}
  \delta(\varphi,c) = \delta(x,c) = \delta(y,c) :=
    \begin{cases}
      \sfE x \lor \sfE y &\text{if } c = a\,, \\
      \sfE y             &\text{if } c \neq a\,.
    \end{cases}
\end{align*}
\upqed
\end{Exam}

\begin{Prop}\label{Prop: formula -> automaton}
Let $L$~be a family of logics over\/~$\bbS$ that is closed under finite disjunctions,
finite conjunctions, and duals.
\begin{enuma}
\item Every $\mu L$-definable language is recognised by an $L$-automaton.
\item Every $\mup L$-definable language is recognised by a pure $L$-automaton.
\item Every $\muaf L$-definable language is recognised by a weak $L$-automaton.
\end{enuma}
\end{Prop}
\begin{proof}
For technical reasons, we will present our translation for formulae with free fixed-point
variables. Therefore we have to work with trees equipped with additional information specifying
the values of the variables. We will represent such a tree as a tree over
the extended alphabet $\Sigma \times \PSet(V)$, where $V$~is the set of variables.
The labels of such a tree are therefore pairs $\langle a,U\rangle$ with $a \in \Sigma$
and $U \subseteq V$, where the second component specifies to which of the variables $x \in V$
the current vertex belongs. Our notation for such trees is
\begin{align*}
  t,\bar P \in \bbT_\bbS[\Sigma \times \PSet(V)]
  \qtextq{where}
  t \in \bbT_\bbS\Sigma \text{ and } P_x \subseteq \dom(t), \text{ for } x \in V\,.
\end{align*}

Fix a formula $\varphi \in \mu L[\Sigma;X_\rmf,Y_\rmf]$.
By Lemma~\ref{Lem: normal form for mu L}, we may assume that $\varphi$~is guarded and
in negation normal form.
For each fixed-point variable~$x$ bound in~$\varphi$, we denote by $\sigma_x x.\psi_x$
the subformula where~$x$ is bound.
Set
\begin{align*}
  X_\rmb &:= \set{ x }{ \psi_x \in \mu L[\Sigma;U,V] \text{ for some $U,V$ with } x \in U }\,, \\
  Y_\rmb &:= \set{ x }{ \psi_x \in \mu L[\Sigma;U,V] \text{ for some $U,V$ with } x \in V }\,.
\end{align*}
Note that, since $\varphi$~is in negation normal form, we have
$\psi \in L[\Sigma;X_\rmf + X_\rmb,Y_\rmf + Y_\rmb]$, for all subformulae~$\psi$ of~$\varphi$.

We define the \emph{alternation-depth} of a variable~$x$ as the length~$n$ of the
longest sequence $\sigma_0 y_0.\psi_0,\dots,\sigma_n y_n.\psi_n$ of subformulae in~$Q$ such that
\begin{itemize}
\item $\sigma_0 y_0.\psi_0 \in Q$,
\item $\sigma_n y_n.\psi_n = \sigma_x x.\psi_x$, and
\item $\sigma_{i+1} y_{i+1}.\psi_{i+1}$ is a subformula of $\sigma_i y_i.\psi_i$ and
  $\sigma_{i+1} \neq \sigma_i$, for all $i < n$.
\end{itemize}

We construct an automaton~$\calA_\varphi$ whose set of states
\begin{align*}
  Q \subseteq L[\Sigma;X_\rmf + X_\rmb,Y_\rmf + Y_\rmb]
\end{align*}
is the set of all subformulae of~$\varphi$.
(We treat different occurrences of the same subformula as different subformulae.)
Each state $\psi \in Q$ checks whether a subtree satisfies the formula~$\psi$.
We use the second component~$i$ only to signal when we iterate a fixed point.
The initial state is~$\varphi$ and
the priorities are given by
\begin{align*}
  \Omega(\psi) := \begin{cases}
                    2i_x+1 &\text{if } \psi = x \in X_\rmb \cup Y_\rmb \text{ and } \sigma_x = \mu\,, \\
                    2i_x   &\text{if } \psi = x \in X_\rmb \cup Y_\rmb \text{ and } \sigma_x = \nu\,, \\
                    2k     &\text{otherwise}\,,
                  \end{cases}
\end{align*}
where $i_x$~is the alternation-depth of the variable~$x$ and
$k$~is the maximal alternation-depth of a variable in~$\varphi$.

We define the transition function~$\delta$ inductively starting with the letters and
the modal operators.
\begin{alignat*}{-1}
  \delta\bigl(a,\langle c,U\rangle\bigr) &:=
    \begin{cases}
      \mathsf{true}  &\text{if } a = c\,, \\
      \mathsf{false} &\text{if } a \neq c\,, \\
    \end{cases}
    &&\quad\text{for } a \in \Sigma\,, \\
  \delta\bigl(\neg a,\langle c,U\rangle\bigr) &:=
    \begin{cases}
      \mathsf{false} &\text{if } a = c\,, \\
      \mathsf{true}  &\text{if } a \neq c\,, \\
    \end{cases}
    &&\quad\text{for } a \in \Sigma\,, \\
  \delta\bigl(x,\langle c,U\rangle\bigr) &:=
    \begin{cases}
      \mathsf{true}  &\text{if } a \in U\,, \\
      \mathsf{false} &\text{if } a \notin U\,, \\
    \end{cases}
    &&\quad\text{for } x \in X_\rmf \cup Y_\rmf\,, \displaybreak[0]\\
  \delta\bigl(\psi_0\lor\psi_1,\langle c,U\rangle\bigr) &:=
    \delta\bigl(\psi_0,\langle c,U\rangle\bigr) \lor
    \delta\bigl(\psi_1,\langle c,U\rangle\bigr)\,, \\
  \delta\bigl(\psi_0\land\psi_1,\langle c,U\rangle\bigr) &:=
    \delta\bigl(\psi_0,\langle c,U\rangle\bigr) \land
    \delta\bigl(\psi_1,\langle c,U\rangle\bigr)\,, \\
  \delta\bigl(\mu x.\psi_0,\langle c,U\rangle\bigr) &:=
    \delta\bigl(\psi_0,\langle c,U\rangle\bigr)\,, \\
  \delta\bigl(\nu x.\psi_0,\langle c,U\rangle\bigr) &:=
    \delta\bigl(\psi_0,\langle c,U\rangle\bigr)\,, \\
  \delta\bigl({\medcircle}\psi,\langle c,U\rangle\bigr) &:= \psi\,, \\
  \delta(x,\langle c,U\rangle) &:= \delta(\psi_x,\langle c,U\rangle)\,,
    &&\quad\text{for } x \in X_\rmb \cup Y_\rmb\,.
\end{alignat*}
(Note that, since $\varphi$~is guarded, the definition of $\delta(\psi_x,\langle c,U\rangle)$
does not depend on the definition of $\delta(x,\langle c,U\rangle)$.
So the above definition does not create a cyclic dependency.)

Before proving that the resulting automaton~$\calA_\varphi$ recognises the correct language,
let us show that it is well-formed and that is has the correct type.
Fix a component~$C$ of~$\calA_\varphi$.
Set
\begin{align*}
  Z := \set{ x \in X_\rmb \cup Y_\rmb }{ \sigma_x x.\psi_x \in C }\,.
\end{align*}
Then
\begin{align*}
  C \subseteq \mu L[\Sigma;X_\rmf + (Z \cap X_\rmb),Y_\rmf + (Z \cap Y_\rmb)\,.
\end{align*}
Let $\Phi_+ \subseteq C$ be the set of all subformulae of~$\varphi$ that contain some variable
from $Z \cap X_\rmb$ and let $\Phi_- \subseteq C$ the the corresponding set for variables
in $Z \cap Y_\rmb$.
It follows that
\begin{align*}
  {\medcircle}\psi \in C
  \qtextq{implies}
  \psi \in L[\mu L,\Phi_+,\Phi_-]\,.
\end{align*}
Since every transition formula is a boolean combination of such formulae~$\psi$, it follows that
\begin{align*}
  \delta\bigl(\psi,\langle c,U\rangle\bigr) \in L[Q,\Phi_+,\Phi_-]\,,
  \quad\text{for all } \psi \in C\,.
\end{align*}
Consequently, $\calA_\varphi$~is well-formed.

Furthermore, if $\varphi$~is pure, we have $Z \subseteq X_\rmb$ or $Z \subseteq Y_\rmb$.
Consequently, $\Phi_+ = \emptyset$ or $\Phi_- = \emptyset$, and it follows that
$\calA_\varphi$~is pure.

In order to obtain a weak automaton we have to slightly modify the above construction.
Note that, if our formula~$\varphi$ is weak, we have
$\sigma_x = \sigma_y$, for all variables $x,y \in C$.
Consequently, all such variables~$x$ have the same alternation depth
and therefore the same priority~$\Omega(x)$.
Since every loop contains a state $x \in X_\rmb \cup Y_\rmb$ and
the other states have maximal priority, it follows that we can set all priorities in~$C$
to the same value~$\Omega(x)$ without changing the behaviour of the automaton.
The resulting automaton is weak.

To prove the correctness of our construction,
let $\calA_\psi$~be the automaton obtained by translating the formula $\psi \in Q$.
Note that each of these automata is equal to part of the automaton~$\calA_\varphi$ for the
whole formula~$\varphi$, except that the transition function differs for states of the form
$\langle x,i\rangle$ with $x \in X_\rmb + Y_\rmb$.

We show by induction on $\psi \in Q$ that
\begin{align*}
  t,\bar P \in \lsem\calA_\psi\rsem
  \quad\iff\quad
  t \in \lsem\psi\rsem_{\bar P}
\end{align*}
(where $\bar P$~are the value of the free variables in~$\psi$).
Most cases are straightforward.
Let us give the proof for the least fixed-point operator.
Hence, let us consider a subformula $\mu x.\psi \in Q$ and
suppose that we have already proved the claim for the formula~$\psi$.

$(\Rightarrow)$
Fix a tree $t,\bar P \in \bbT_\bbS(\Sigma \times \PSet(X_\rmf + Y_\rmf))$.
Suppose that $\sigma$~is a winning strategy for Player~$\Diamond$ in the game
$\calG(\calA_{\mu x.\psi},t,\bar P)$. Since no infinite play conforming to~$\sigma$ contains
infinitely many positions with the state~$x$, we can define the following ordinal rank for
the positions in the game. If, from a postion $\langle v,q\rangle$, no position of the form
$\langle u,x\rangle$ is reachable when following the strategy~$\sigma$, we assign the rank~$0$
to $\langle v,q\rangle$. Inductively, the rank of an arbitrary position $\langle v,q\rangle$
of Player~$\Diamond$ is the least ordinal~$\alpha$ such that every successor of the position
$\sigma(\langle v,q\rangle)$ has a rank less than~$\alpha$.

By induction on~$\alpha$ we prove that, if Player~$\Diamond$ has a winning strategy~$\sigma$
of rank at most~$\alpha$ in the game $\calG(\calA_{\mu x.\psi},t,\bar P)$, then
$t \in \lsem\mu x.\psi\rsem_{\bar P}$. Let $Q$~be the set of all vertices $v \in \dom(t)$
such that the position $\langle v,x\rangle$ of $\calG(\calA_{\mu x.\psi},t,\bar P)$ has
rank less than~$\alpha$. Then $\sigma$~induces a winning strategy
in~$\calG(\calA_\psi,t,\bar PQ)$, which implies that $t \in \lsem\psi\rsem_{\bar P,Q}$.
By inductive hypothesis, we further have $Q \subseteq \lsem\mu x.\psi\rsem_{\bar P}$.
By monotonicity of~$\psi$, it follows that $t \in \lsem\psi(\mu x.\psi)\rsem_{\bar P}$,
which is equivalent to $t \in \lsem\mu x.\psi\rsem_{\bar P}$,

$(\Leftarrow)$
Fix a tree $t,\bar P \in \bbT_\bbS(\Sigma \times \PSet(X_\rmf + Y_\rmf))$.
For an ordinal~$\alpha$, let $F^\alpha := F_\psi^\alpha(\emptyset)$ be the $\alpha$-th
stage of the fixed-point induction for the formula~$\psi$ on~$t,\bar P$.
By induction on~$\alpha$, we show that $v \in F^\alpha$ implies
$(t,\bar P)|_v \in \lsem\calA_{\mu x.\psi}\rsem$,
where $(t,\bar P)|_v$~denotes the subtree of~$t,\bar P$ rooted at~$v$.
For $\alpha = 0$, the claim is trivial.
If $\alpha$~is a limit ordinal, we have $F^\alpha = \bigcup_{\beta<\alpha} F^\beta$
and the claim follows immediately by inductive hypothesis.
For the successor step, suppose that we have already proved the claim for~$\alpha$.
Let $v \in F^{\alpha+1}$.
By the inductive hypothesis for~$\psi$, there exists an accepting run of~$\calA_\psi$
on~$t,\bar PF^\alpha$.
Furthermore, for every $u \in F^\alpha$, the inductive hypothesis for~$\alpha$ provides a run
of~$\calA_{\mu x.\psi}$ on~$(t,\bar P)|_u$. Combining these runs, we obtain a run
of~$\calA_{\mu x.\psi}$ on~$(t,\bar P)|_v$.
\end{proof}

\section{Projection}   
\label{Sect: projection}

We would like to translate various variants of monadic second-order logic into automata.
To be able to do so, we need to prove that the resulting classes of
automata are closed under projections of various kinds.
\begin{Def}\label{Def: projection}
Let $\bbF$~be a polynomial functor and $\Sigma,\Gamma$ two sets.

(a)
The \emph{projection} of $s \in \bbF(\Sigma \times \Gamma)$ is
\begin{align*}
  \pr_\Sigma(s) := f \circ s\,,
\end{align*}
where $f : \Sigma \times \Gamma \to \Sigma$ is the projection to the first component.

The \emph{projection} of a language $K \subseteq \bbF(\Sigma \times \Gamma)$
is the language
\begin{align*}
  \pr_\Sigma[K] := \set{ \pr_\Sigma(s) }{ s \in K }\,.
\end{align*}

(b)
Fix a distinguished element $\gamma_0 \in \Gamma$ and a class
\begin{align*}
  \calP \subseteq
    \bigset{ P }{ P \subseteq \dom(t), \text{ for some } t \in \bbF\one }\,.
\end{align*}
(Usually, $\Gamma = \PSet(X)$ is a power set and $\gamma_0 = \emptyset$ the empty set.)

The \emph{$\calP$-projection} of $K \subseteq \bbT(\Sigma \times \Gamma)$ is the language
$\pr^\calP_{\Sigma,\gamma_0}[K] := \pr_{\Sigma,\gamma_0}[K_\calP]$,
where
\begin{align*}
  K_\calP :=
    \bigset{ s \in K }
           { \text{there is some } P \in \calP \text{ such that }
             \dom(s) \setminus t^{-1}[\Sigma \times \{\gamma_0\}] \subseteq P }\,.
\end{align*}

If $\bbF = \bbT_\bbS$ and the set~$\calP$ consists of all
\emph{finite sets,} \emph{well-founded sets,}
\emph{finitely branching sets,} \emph{chains,} or \emph{finite chains,} we speak of,
respectively, the \emph{finite projection,} the \emph{well-founded projection,}
the \emph{finitely branching projection,} the \emph{chain projection,} or the
\emph{finite chain projection} of~$K$.

If $\bbF = \bbS$ and the set~$\calP$ consists of all
\emph{finite sets} or all \emph{singletons,} we speak of, respectively, the
\emph{finite projection} or the \emph{singleton projection.}

(c) We say that a property~$P$ of languages is \emph{closed under $\calP$-projections}
if, whenever a language~$K$ has property~$P$, so does every $\calP$-projection of~$K$.
Similarly, we say that a family of logics~$L$ is closed under $\calP$-projections if,
whenever a language~$K$ is $L$-definable, so is every $\calP$-projection of~$K$.
Given a formula $\varphi \in L[\Sigma \times \Gamma]$ defining~$K$,
we denote by $\exists^\calP_{\Gamma,\gamma_0}\varphi$
the formula defining the corresponding $\calP$-projection.
\end{Def}

The automaton construction below does not work for arbitrary logics~$L$.
We have to make a few restrictions. First of all, the construction does not respect polarities.
Which means that we will work with families of logics \emph{without} polarities.
Furthermore, we require two basic closure properties.
\begin{Def}
Let $L$~be a family of logics over $\bbS\circ\PSet$.

(a)
We say that $L$~is closed under \emph{label restrictions} if,
for all $\Delta \subseteq \PSet(\Sigma)$,
there exists an $L[\Sigma]$-formula~$\chi_\Delta$ defining the set~$\bbS\Delta$.

(b)
$L$~is closed under \emph{boolean label substitutions}
if there exists a family~$\hat L$ of logics over~$\bbS$ such that
$L[Q] = \hat L[\PSet(Q)]$, for all sets~$Q$.
\end{Def}
\begin{Lem}\label{Lem: boolean substitutions}
Let $L$~be a family of logics over $\bbS \circ \PSet$ that is closed under boolean
label substitutions, Let $\Sigma,\Gamma,\Delta$ be sets, $\psi \in L[\Sigma]$ a formula,
and let $(\vartheta_c)_{c \in \Sigma}$ be a family of boolean combinations of elements of
$\Delta \times \Gamma$. There exists an $L[\Delta \times \PSet(\Gamma)]$-formula
$\psi[c \mapsto \vartheta_c]_{c \in \Sigma}$ such that
\begin{align*}
  s \models \psi[c \mapsto \vartheta_c]_{c \in \Sigma}
  \quad\iff\quad
  \bbS\tau(s) \models \psi\,,
\end{align*}
where $\tau : \PSet(\Delta \times \PSet(\Gamma)) \to \PSet(\Sigma)$ is the function
\begin{align*}
  \tau(P) := \set{ c \in \Sigma }
                 { \langle a,B\rangle \in P,\ \{a\} \times B \text{ satisfies } \vartheta_c }\,.
\end{align*}
\end{Lem}
\begin{proof}
We have to show that $\bbS\tau : \bbS\PSet(\Delta \times \PSet(\Gamma)) \to \bbS\PSet(\Sigma)$
is (the second component of) a morphism $L[\Sigma] \to L[\Delta \times \PSet(\Gamma)]$ of logics.
Note that
\begin{align*}
  \tau = g \circ \PSet(f)\,,
\end{align*}
where
\begin{align*}
  f &: \Delta \times \PSet(\Gamma) \to \PSet(\Delta \times \Gamma)
    : \langle a,B\rangle \mapsto \{a\} \times B\,, \\
  g &: \PSet\PSet(\Delta \times \Gamma) \to \PSet(\Sigma)
    : P \mapsto \set{ c \in \Sigma }{ A \in P,\ A \text{ satisfies } \vartheta_c }\,.
\end{align*}
Let $\hat L$~be the extension of~$L$ to a logic over~$\bbS$.
We obtain morphisms of logics
\begin{align*}
  \bbS\PSet(f) &: L[\PSet(\Delta \times \Gamma)] \to L[\Delta \times \PSet(\Gamma)]\,, \\
  \bbS g &: \hat L[\Sigma] \to \hat L[\PSet(\Delta \times \Gamma)]\,.
\end{align*}
The composition $\bbS\tau = \bbS(g \circ \PSet(f))$ induces the desired morphism of logics
$L[\Sigma] \to L[\Delta \times \PSet(\Gamma)]$.
\end{proof}

\begin{Def}
Let $\calA = \langle Q,\Sigma,\delta,q_0,\Omega\rangle$ be an $L$-automaton.

(a) For $s,t \in \bbS\PSet(X)$, we write
\begin{align*}
  s \leq t \quad\defiff\quad
  s \simeq_\sh t \qtextq{and} s(d) \subseteq t(d)\,, \quad\text{for all } d \in \dom(s)\,.
\end{align*}

(b)
$\calA$ is \emph{partially non-deterministic} if there exists a partition
$Q = Q_{\mathrm{alt}} + Q_{\mathrm{nd}}$ of its states such that,
\begin{itemize}
\item for $q \in Q_{\mathrm{alt}}$, the formula $\delta(q,a)$ does not contain any state
  from~$Q_{\mathrm{nd}}$,
\item for all $q \in Q_{\mathrm{nd}}$ and $a \in \Sigma$,
  \begin{align*}
    s \models \delta(q,a)
    \qtextq{implies}
    &\text{there is some } s_0 \leq s \text{ such that}\\
    &s_0 \models \delta(q,a) \text{ and, for all } d \in \dom(s) \\
    &\abs{s_0(d)} = 1 \text{ or } s_0(d) \subseteq Q_{\mathrm{alt}}\,.
  \end{align*}
\end{itemize}
We call the elements of~$Q_{\mathrm{alt}}$ the \emph{alternating states} of~$\calA$
and those of~$Q_{\mathrm{nd}}$ its \emph{non-deterministic states.}

(c) Suppose that $\calA$~is partially non-deterministic.
Let $Q = Q_{\mathrm{alt}} + Q_{\mathrm{nd}}$ be the corresponding partition of its states,
and let $\rho$~be a run on the input tree~$t$.

We say that $\rho$~is \emph{economical} if, for all $v \in \dom(\rho)$,
\begin{align*}
  \abs{\rho(v)} = 1
  \qtextq{or}
  \rho(v) \subseteq Q \times Q_{\mathrm{alt}}\,.
\end{align*}
The \emph{non-deterministic part} of an economical run~$\rho$ on~$t$ is the set
\begin{align*}
  P := \set{ v \in \dom(\rho) }{ \rho(v) \in \PSet(Q_{\mathrm{nd}} \times Q_{\mathrm{nd}}) }\,.
\end{align*}
The complement of~$P$ is the \emph{alternating part} of~$\rho$.

Given a class~$\calP$ of prefixes as in Definition~\ref{Def: projection}, we say that $\calA$~is
\emph{partially non-deterministic of shape~$\calP$} if, for every $t \in \lsem\calA\rsem$,
there exists an accpeting economical run~$\rho$ whose non-deterministic part belongs to~$\calP$.
\end{Def}
\begin{Lem}
Let $\calA$~be a partially non-deterministic automaton. For every $t \in \lsem\calA\rsem$,
there exists an accpeting run~$\rho$ of~$\calA$ on~$t$ that is economical.
\end{Lem}
\begin{proof}
Let $\rho$~be a minimal (with respect to~$\subseteq$) accepting run of~$\calA$ on~$t$.
We claim that~$\rho$ is economical.
For a contradiction, suppose otherwise. Fix a vertex $v \in \dom(t)$ such that
\begin{align*}
  \abs{\rho(v)} > 1 \qtextq{and} \rho(v) \cap (Q \times Q_{\mathrm{nd}}) \neq \emptyset\,.
\end{align*}
Choose~$v$ such that $\abs{v}$ is minimal, let $u$~be the predecessor of~$v$, and fix
a pair $\langle p,q\rangle \in \rho(v)$ with $q \in Q_{\mathrm{nd}}$.

We distinguish two cases.
If $p \in Q_{\mathrm{alt}}$, the transition formula $\delta(p,t(u))$ does not mention
the state $q \in Q_{\mathrm{nd}}$. Hence, we can remove the pair $\langle p,q\rangle$
from~$\rho(v)$ and still obtain a valid run. A~contradiction to the minimality of~$\rho$.

Hence, we have $p \in Q_{\mathrm{nd}}$.
Since $\calA$~is partiall non-deterministic, it follows that there exists some
$s \simeq_\sh \suc(u)$ with
\begin{itemize}
\item $s \models \delta(p,t(u))$,
\item $\abs{s(d)} = 1$ or $s(d) \subseteq Q_{\mathrm{alt}}$,
\item $q' \in s(d) \Rightarrow \langle p,q'\rangle \in \rho(\suc(u)(d))$.
\end{itemize}
Let $d$~be the direction such that $v = \suc(u)(d)$.
If $s(d) = \{q'\}$, we could remove everything from~$\rho(v)$ except for the pair
$\langle p,q'\rangle$.
If $s(d) \subseteq Q_{\mathrm{alt}}$, we could remove the pair $\langle p,q\rangle$
from~$\rho(v)$. In both cases we obtain a contradiction to the minimality of~$\rho$.
\end{proof}

By definition, every partially non-deterministic $L$-automaton is an alternating $L$-automaton.
Conversely, we can translate every alternating $L$-automaton into a partially non-deterministic
one.
To do so, we make use of the following consequence of the theorem of
McNaughton-Pappert~\cite{McNaughton66}.
\begin{Prop}\label{Prop: deterministic automaton for omega-semigroup}
For every finite $\omega$-semigroup\/~$\frakS = \langle S,S_\omega\rangle$ and every element
$a \in S_\omega$, there exists a deterministic $\omega$-automaton~$\calA$ such that
\begin{align*}
  \lsem\calA\rsem = \set{ w \in S^\omega }{ \pi(w) = a }\,.
\end{align*}
\end{Prop}

\begin{Prop}\label{Prop: alternating => non-deterministic}
Let $T \in \{\mathrm{general},\mathrm{pure},\mathrm{weak}\}$ be a type of automaton,
let $L$~be a family of logics without\?(!) polarities for $\bbS \circ \PSet$
that is closed under conjunctions, disjunctions, label restrictions, and boolean label
substitutions, and let $L' \in \{L,L_\rmc,L_\rmd\}$.
Suppose that $L$~and~$L'$ satisfy one of the following conditions.
\begin{itemize}
\item $L' = L$ and $L$~is closed under projections.
\item $L' = L_\rmc$ and $L$~is closed under finite projections.
\item $L' = L_\rmd$ and $L$~is closed under singleton projections.
\end{itemize}

For every alternating $L'$-automaton~$\calA$ of type~$T$, there exists an
$L'$-automaton~$\calB$ of type~$T$ such that $\lsem\calB\rsem = \lsem\calA\rsem$ and
$\calB$~is partially non-deterministic whose shape is given by the following table.

\medskip
\noindent\centering
\begin{tabular}{c@{\hskip1.5em}lll}
\toprule
  type     & general                  & pure                     & weak \\
\midrule
  $L$      & all trees                & all trees                & well-founded trees \\
  $L_\rmc$ & finitely-branching trees & finitely-branching trees & finite trees \\
  $L_\rmd$ & chains                   & chains                   & finite chains \\
\bottomrule
\end{tabular}
\end{Prop}
\begin{proof}
We start with the cases where $T = \mathrm{general}$ or $T = \mathrm{pure}$,
and then we explain how to modify this proof for weak automata.
Let $\calA = \langle Q,\Sigma,\delta,q_0,\Omega\rangle$ be an alternating $L'$-automaton,
possibly pure.
We construct a partially non-deterministic automaton~$\calB$ whose alternating part is
just~$\calA$, while the non-deterministic part guesses a run~$\rho$ of~$\calA$ and
verifies that it is accepting. To do so, $\calB$~has to check that every trace of~$\rho$
satisfies the parity condition. Unfortunately, this condition is not a parity condition
itself. Hence, we first have to construct a deterministic $\omega$-automaton~$\calC$
that reads a branch of~$\rho$ and checks all the traces along that branch.
Then we simply have to run $\calC$~along all the branches of~$\rho$ and
use the states of~$\calC$ for our parity condition.

The automaton~$\calC$ is based on the $\omega$-semigroup $\frakS = \langle S,S_\omega\rangle$
with domains
\begin{align*}
  S := \PSet(Q \times Q)
  \qtextq{and}
  S_\omega := \PSet(Q)\,,
\end{align*}
where the product is defined as follows.
For $A,B,A_0,A_1,\ldots \in S$ and $U \in S_\omega$, we set
\begin{align*}
  A\cdot B &:= \bigset{ \langle p,r\rangle }
                      { \langle p,q\rangle \in A \text{ and } \langle q,r\rangle \in B }\,, \\
  A\cdot U &:= \bigset{ p }{ \langle p,q\rangle \in A \text{ and } q \in U }\,, \\
  \prod_{i<\omega} A_i &:=
    \begin{aligned}[t]
      \biglset p_0 \in Q \bigmset {}
        & \text{there are } p_0,p_1,\dots \in Q \text{ with }
          \langle p_i,p_{i+1}\rangle \in A_i, \text{ for all } i, \\
        & \text{such that } \liminf\nolimits_i \Omega(p_i) \text{ is even} \bigrset\,.
    \end{aligned}
\end{align*}
By Proposition~\ref{Prop: deterministic automaton for omega-semigroup}
there exists a deterministic $\omega$-automaton
$\calC = \langle W,S,\eta,w_0,\Phi\rangle$
recognising the language
\begin{align*}
  \lsem\calC\rsem = \set{ (A_i)_i \in S^\omega }{ q_0 \in \textstyle\prod_i A_i }\,.
\end{align*}

We construct a partially non-deterministic automaton
$\calB := \langle Q',\Sigma,\delta',q'_0,\Omega'\rangle$ as follows.
For the alternating part, we use the original states of~$\calA$, while
the non-deterministic part uses states in $W \times S$. For technical reasons,
when switching from non-deterministic to alternating mode, the automaton goes through
a state in $Q \times Q$ (the first component contains the previous state and the second
component the current one). Hence, we set
\begin{align*}
  Q'_{\mathrm{alt}} := Q + Q \times Q
  \qtextq{and}
  Q'_{\mathrm{nd}} := W \times S\,.
\end{align*}
Initial state and priority function are given by
\begin{align*}
  q'_0 &:= \bigl\langle w_0,\, \{\langle q_0,q_0\rangle\}\bigr\rangle\,, \\
  \Omega'(p) &:=
    \begin{cases}
      \Omega(p)  &\text{if } p \in Q\,, \\
      \Omega(q') &\text{if } p = \langle p',q'\rangle \in Q \times Q\,, \\
      \Phi(w)    &\text{if } p = \langle w,A\rangle \in W \times S\,,
    \end{cases}
\end{align*}
Finally, the transition function is defined as follows. For the alternating part, we set
\begin{align*}
  \delta'(p,a) &:= \delta(p,a)\,, \qquad\text{for } p \in Q\,, \\
  \delta'(\langle p,q\rangle,a) &:= \delta(q,a)\,,
    \qquad\text{for } \langle p,q\rangle \in Q \times Q\,.
\end{align*}

To define the transition function for the non-deterministic part, we have to deal with
the problem that a (non-economical) run might assign several non-deterministic states
to the same vertex. What we do in this case is to guess one of these states and ignore the
others. Hence, given a labelling $s \in \bbS\PSet(Q')$, we use an inverse projection
to guess a labelling $s' \in \bbS\Delta$ where
\begin{align*}
  \Delta :=
    \bigset{ \langle P,P^\dag\rangle \in \PSet(Q') \times \PSet(Q'_{\mathrm{nd}}) }
           { P^\dag \subseteq P,\ \abs{P^\dag} \leq 1 }\,,
\end{align*}
and then we ignore all non-deterministic states that do not belong to~$P_0$.
To distinguish given states in $p \in P \cap Q'_{\mathrm{nd}}$ from the guessed ones
in~$P^\dag$, we denote the latter states by~$p^\dag$. Hence, the formula~$p^\dag$ checks
whether $p \in P^\dag$ while $p$~checks whether $p \in P$.
This leads to the transition formula
\begin{align*}
  \delta'(\langle w,A\rangle,a) :=
    \exists^\calP_{\PSet(Q'),\emptyset}
      \Bigl[\chi_\Delta \land \Land_{\langle p',p\rangle \in A}\hat\delta_p\Bigr]
\end{align*}
where $\chi_\Delta \in L[\PSet(Q') \times \PSet(Q'_{\mathrm{nd}})]$ is a formula stating that
all labels belong~$\Delta$ (which exists since $L$~is closed under label restrictions), and
\begin{align*}
  \calP &:= \begin{cases}
              \text{all sets}    &\text{if } L' = L\,, \\
              \text{finite sets} &\text{if } L' = L_\rmc\,, \\
              \text{singletons}  &\text{if } L' = L_\rmd\,,
            \end{cases} \\
  \hat\delta_p &:=
    \delta(p,a)\Bigl[q \mapsto (\vartheta \land \langle p,q\rangle^\dag)
                          \lor\!\! \Lor_{B \owns \langle p,q\rangle} \langle \eta(w,A),B\rangle^\dag
                     \Bigr]_{q \in Q}\,, \\
  \vartheta &:= \Land_{q \in Q'_{\mathrm{nd}}} \neg q^\dag\,.
\end{align*}
($\delta(p,a)[q \mapsto \vartheta_q]_{q \in Q}$ is the formula from
Lemma~\ref{Lem: boolean substitutions}.)

The automaton~$\calB$ is a well-formed $L'$-automaton since every non-deterministic state
occurs positively in the formula $\delta'(\langle w,A\rangle,a)$ and
\begin{align*}
  \delta'(\langle w,A\rangle,a) \in L'[Q',Q'_{\mathrm{nd}},\emptyset]\,.
\end{align*}
(We can remove all non-deterministic states that are different from the guessed ones.
And there are only finitely many guessed states if $L' = L_\rmc$,
and only a single one if $L' = L_\rmd$.)
Furthermore it follows that, if $\calA$~is pure, then so is~$\calB$.

Let us check that $\calB$~is partially non-deterministic with the desired shape
and with alternating states $Q'_{\mathrm{alt}} = Q + Q \times Q$ and non-deterministic states
$Q'_{\mathrm{nd}} = W \times S$.
To check that $\calB$~is partially non-deterministic, suppose that
\begin{align*}
  s \models \delta'(q,a)\,, \quad\text{for } q \in Q'_{\mathrm{nd}} \text{ and } a \in \Sigma\,.
\end{align*}
By definition of the transition formula for non-deterministic states, it follows that there
is some $s_0 \leq s$ such that
\begin{align*}
  s_0 \models \delta'(q,a) \text{ and, for all } d \in \dom(s),\
  \abs{s_0(d)} = 1 \text{ or } s_0(d) \subseteq Q'_{\mathrm{alt}}\,.
\end{align*}
The fact that the shape of~$\calB$ is either the class of all trees, of all finitely-branching
ones, or of all chains, follows by choice of~$\calP$ in the transition formula above.

It remains to prove that $\lsem\calB\rsem = \lsem\calA\rsem$.
First, note that we have
\begin{align*}
  s \models \delta'(\langle w,A\rangle,a)
\end{align*}
if, and only if, there exists some $s_0 \leq s$ such that
\begin{itemize}
\item $\abs{s_0(d)} = 1 \text{ or } s_0(d) \subseteq Q \times Q\,,
  \quad\text{for all } d \in \dom(s)\,,$
\item $\set{ d \in \dom(s) }{ s_0(d) \in Q_{\mathrm{nd}} } \in \calP$
\item $\bbS\tau_p(s_0) \models \delta(p,a)\,, \quad\text{for all } \langle p',p\rangle \in A\,,$
  where
  \begin{align*}
    \tau_p(P) :=
      \begin{cases}
        f_p[P]\quad{} &\text{if } P \subseteq Q \times Q\,, \\[0.5em]
        \rlap{\set{ q \in Q }
                  { \langle w',B\rangle \in P,\ w' = \eta(w,A) \text{ and }
                    \langle p,q\rangle \in B }} \\
          &\text{if } P \subseteq W \times S\,.
      \end{cases}
  \end{align*}
  and $f_p(B) := \set{ q \in Q }{ \langle p,q\rangle \in B }$.
\end{itemize}
It follows that $s$~is a model of $\delta'(\langle w,A\rangle,a)$ if, and only if,
there is some $s' \in \bbS[\PSet(Q \times Q)]$ such that
\begin{align*}
  s'(d) \subseteq s(d) \cap Q_{\mathrm{alt}}
  \qtextq{or}
  \langle\eta(w,A),\, s'(d)\rangle \in s(d)\,,
\end{align*}
and
\begin{align*}
  \bbS f_p(s') \models \delta(p,a)\,, \quad\text{for all } \langle p',p\rangle \in A\,,
\end{align*}
where $f_p : Q \times Q \to Q$ is defined as above.

$(\subseteq)$ Let $\rho' : \dom(t) \to \PSet(Q' \times Q')$ be an accepting economical run
of~$\calB$ on some tree~$t$.
We define a function $\rho : \dom(t) \to \PSet(Q \times Q)$ by
\begin{align*}
  \rho(v) :=
    \begin{cases}
      \rho'(v) &\text{if } \rho'(v) \subseteq Q \times Q\,, \\
      \bigset{ \langle p',q\rangle }{ \langle\langle p,p'\rangle,q\rangle \in \rho'(v) }
               &\text{if } \rho'(v) \subseteq (Q \times Q) \times Q\,, \\[0.25em]
      B        &\mkern-144mu\text{if } \rho'(v) = \{\langle w,A\rangle\} \times B
                            \text{ for some } B \subseteq Q \times Q\,, \\
      B        &\mkern-144mu\text{if } \rho'(v) = \{\langle w,A,u,B\rangle\}
                                       \subseteq (W \times S) \times (W \times S)\,,
    \end{cases}
\end{align*}
By the above remark, it follows that $\rho$~is an accepting run of~$\calA$ on~$t$.

$(\supseteq)$
Let $\rho : \dom(t) \to \PSet(Q \times Q)$ be an accepting run of~$\calA$ on some tree~$t$.
W.l.o.g.\ we may assume that, for every vertex $v \in \dom(t)$ with predecessor~$u$
and every pair $\langle q,q'\rangle \in \rho(v)$, there is some $\langle p,p'\rangle \in \rho(u)$
with $p' = q$. (Otherwise, we can remove from~$\rho(v)$ all pairs $\langle q,q'\rangle$
not satisfying this condition.)
Given a prefix-closed subset $P \subseteq \dom(t)$, we will construct a run
$\rho' : \dom(t) \to \PSet(Q' \times Q')$ whose non-deterministic part is~$P$.
Let $\sigma : \dom(t) \to W$ be the function with
\begin{align*}
  \sigma(v) :=
    \begin{cases}
      w_0 &\text{if } v = \emptyseq \text{ is the root,} \\
      \eta(\sigma(u),\rho(u)) &\text{if } v \text{ has a predecessor } u\,.
    \end{cases}
\end{align*}
We define $\rho' : \dom(t) \to \PSet(Q' \times Q')$ as follows.
For a vertex $v \in \dom(t)$ with predecessor~$u$, we set
\begin{align*}
  \rho'(v) :=
    \begin{cases}
      \bigl\{\langle\sigma(u),\rho(u),\sigma(v),\rho(v)\rangle\bigr\} &\text{if } v \in P\,, \\
      \{\langle\sigma(u),\rho(u)\rangle\} \times \rho(v)
              &\text{if } v \notin P \text{ but } u \in P\,, \\
      \rho(v) &\text{if } u \notin P\,.
    \end{cases}
\end{align*}
By the above remark, it follows that $\rho'$~is an accpeting economical run of~$\calB$ on~$t$
whose non-deterministic part is equal to~$P$.

\smallskip
It remains to consider the weak cases. Let $\calB'$~be the automaton obtained from
the automaton~$\calB$ constructed above by increasing the priorities of all states
in~$Q'_{\mathrm{at}}$ by~$2$ and
setting the priorities of all states in~$Q'_{\mathrm{nd}}$ to~$1$.
Since no accpeting run of~$\calB'$ has an infinite branch with non-deterministic states only,
it follows that the shape of~$\calB'$ consists of all well-founded trees in the shape of~$\calB$.
Furthermore, the automaton~$\calB'$ is weak since, as we remarked above,
\begin{align*}
  \delta'(\langle w,A\rangle,a) \in L'[Q',Q'_{\mathrm{nd}},\emptyset]\,.
\end{align*}
\upqed
\end{proof}

\begin{Thm}\label{Thm: closure under projections}
Let $T \in \{\mathrm{general},\mathrm{pure},\mathrm{weak}\}$ be a type of automaton,
let $L$~be a family of logics without\?(!) polarities for $\bbS \circ \PSet$
that is closed under conjunctions, disjunctions, label restrictions, and boolean label
substitutions, and let $L' \in \{L,L_\rmc,L_\rmd\}$.
Suppose that $L$~and~$L'$ satisfy one of the following conditions.
\begin{itemize}
\item $L' = L$ and $L$~is closed under projections.
\item $L' = L_\rmc$ and $L$~is closed under finite projections.
\item $L' = L_\rmd$ and $L$~is closed under singleton projections.
\end{itemize}

The class of languages recognised by $L'$-automata of type~$T$ is closed
under $\calP$-projections where $\calP$~is the class of trees from the following table.

\medskip
\noindent\centering
\begin{tabular}{c@{\hskip1.5em}lll}
\toprule
  type     & general                  & pure                     & weak \\
\midrule
  $L$      & all trees                & all trees                & well-founded trees \\
  $L_\rmc$ & finitely-branching trees & finitely-branching trees & finite trees \\
  $L_\rmd$ & chains                   & chains                   & finite chains \\
\bottomrule
\end{tabular}
\end{Thm}
\begin{proof}
Let $K \subseteq \bbT_\bbS[\Sigma \times \Gamma]$ be a language of the form
$K = \lsem\calA\rsem$ for some $L'$-automaton~$\calA$ of type~$T$
and fix a distinguished element $\gamma_0 \in \Gamma$.
By Proposition~\ref{Prop: alternating => non-deterministic}, there exists a
partially non-deterministic $L'$-automaton
$\calB = \langle Q,\Sigma \times \Gamma,\delta,q_0,\Omega\rangle$
such that $\lsem\calB\rsem = K$ and the shape of~$\calB$ is equal to~$\calP$.
Let $\calC$~be the automaton obtained from~$\calB$ by changing the transition relation to
\begin{align*}
  \delta'(q,a) :=
    \begin{cases}
      \delta(q,\langle a,\gamma_0\rangle)
        &\text{if } q \in Q_{\mathrm{alt}}\,, \\
      \displaystyle
      \Lor_{\gamma \in \Gamma} \delta(q,\langle a,\gamma\rangle)
        &\text{if } q \in Q_{\mathrm{nd}}\,,
    \end{cases}
  \qquad\text{for } q \in Q \text{ and } a \in \Sigma\,.
\end{align*}
We claim that $\lsem\calC\rsem = \pr^\calP_{\Sigma,\gamma_0}[K]$.

$(\supseteq)$ Let $t = \pr^\calP_{\Gamma,\gamma_0}(s)$, for some $s \in K$.
By assumption, there exists an accepting economical run~$\rho$ of~$\calB$ on~$s$
such that the set
\begin{align*}
  P := \set{ v \in \dom(s) }{ s(v) \notin \Sigma \times \{\gamma_0\} }
\end{align*}
is included in the non-deterministic part of~$\rho$.
Then $\rho$~is also an accepting run of~$\calC$ on~$t$.

$(\subseteq)$
Let $\rho$~be an accepting economical run of~$\calC$ on~$t$ and
let $P$~be its non-deterministic part.
Since the shape of~$\calC$ is equal to the shape of~$\calB$, we have $P \in \calP$.
By definition of~$\delta'$ we can choose, for every $v \in P$, some element $c_v \in \Gamma$
such that
\begin{align*}
  s_{v,p} \models \delta(p,\langle t(v),c_v\rangle)\,, \quad\text{for all } v \in P
  \text{ with } \rho(v) = \{\langle q,p\rangle\}\,,
\end{align*}
where $s_{v,p} \in \bbS\PSet(Q)$ is the successor structure obtained from~$\rho$
as in Definition~\ref{Def: automaton}.
We define a tree $s \in \bbT_\bbS(\Sigma \times \Gamma)$ by
\begin{align*}
  s(v) := \begin{cases}
            \langle t(v),\gamma_0\rangle &\text{if } v \notin P\,, \\
            \langle t(v),c_v\rangle      &\text{if } v \in P\,.
          \end{cases}
\end{align*}
Then $t = \pr_{\Sigma,\gamma_0}(s)$ and $\rho$~is an accepting run of~$\calB$ on~$s$.
Hence, $s \in K$ and $t \in \pr_{\Sigma,\gamma_0}[K]$.
\end{proof}

\section{Monadic Second-Order Logic}   
\label{Sect: MSO}

In this section, we use the machinery we have set up to derive characterisations of
certain variants of monadic second-order logic. Let us start by introducing the
logics we will work with.
\begin{Def}
Let $Q$~be an alphabet and $\Sigma$~a signature.

(a) The logic~$\rmE_\infty[Q]$ has formulae that are boolean combination of statements
of the form~$\sfE_k\varphi$ where $k < \omega$ or $k = \infty$, and
$\varphi$~is a boolean combination of elements of~$Q$.
Such a formula holds in a structure $s \in \bbS\PSet(Q)$ if, and only if,
there are at least~$k$ elements $v \in \dom(s)$ such that $s(v)$ satisfies~$\varphi$.

(b) We denote by $\rmE_\omega[Q]$ the fragment of~$\rmE_\infty[Q]$ that only uses the
quantifier $\sfE_k$ with $k < \omega$.
Similarly, $\rmE_1[Q]$ is the fragment of~$\rmE_\infty[Q]$ that only uses the
quantifier $\sfE_k$ with $k =\nobreak 1$.
Finally, we denote by~$\rmC[Q]$ the extension of~$\rmE_\infty[Q]$ by statements of the form
$\sfC_{k,m}\varphi$ stating that the number of elements satisfying~$\varphi$ is finite
and congruent~$k$ modulo~$m$.

(c) For a relational signature~$\Sigma$, we denote by~$\bbS_\Sigma$ the functor mapping
a set~$Q$ to the class of all structures of signature $\Sigma + \set{ P_q }{ q \in Q }$
where the (unary) predicates~$P_q$ form a partition of the universe.

Let $\frakA$~be a structure over a signature of the form
$\Sigma + \{E\} + \set{ P_q }{ q \in Q }$ (where $E$~is binary) such that
the predicates~$P_q$ form a partition of the universe~$A$.
We identify~$\frakA$ with with the $Q$-labelled $\bbS_\Sigma$-enriched transition
system $\langle A,\suc,\lambda\rangle$ where
\begin{align*}
  \lambda(a) := q \quad\defiff\quad a \in P_q\,, \quad\text{for } a \in A \text{ and } q \in Q\,,
\end{align*}
and $\suc(a)$~is the $(\Sigma + \set{ P_q }{ q \in Q })$-reduct of the substructure of~$\frakA$
induced by the set
\begin{align*}
  \set{ b \in A }{ \langle a,b\rangle \in E }\,.
\end{align*}

(d) Let $\calP$~be a property of sets.
We denote by $\MSO_\calP[\Sigma;Q]$ the version of monadic second-order logic where
quantification is restricted to sets included in some set satisfying~$\calP$,
and where the models are structures with the signature $\Sigma + \{E\} + \set{ P_q }{ q \in Q }$.
For particular choices of~$\calP$ we obtain the following variants.

\medskip
{\noindent\centering
\begin{tabular}{cll}
\toprule
  logic   & name & $\calP$ \\
\midrule
  $\MSO$  &monadic second-order logic      & all sets \\
  $\WMSO$ &weak monadic second-order logic & finite sets \\
  $\CL$   &chain logic                     & chains \\
  $\WCL$  &weak chain logic                & finite chains \\
  $\MSO_{\mathrm{wf}}$ &---                & well-founded trees \\
  $\MSO_{\mathrm{fb}}$ &---                & finitely-branching trees \\
\bottomrule
\end{tabular}
\par}

\medskip
(e) \emph{Guarded second-order logic} $\GSO$ is a variant of full second-order logic where
all second-order quantifiers are restricted to range over \emph{guarded} relations only.
A relation~$R$ is called \emph{guarded} if every tuple $\bar a \in R$ is included (as a set)
in some tuple~$\bar c$ from a relation of the given structure. (We consider equality as a
binary relation in this context, which implies that every unary relation is guarded.)

(f) Finally, the \emph{counting} variants of $\MSO$ and $\GSO$ are the extensions
of the respective logic by predicates of the form
\begin{align*}
  \abs{X} < \infty \,\land\, \abs{X} \equiv k \pmod m\,,
  \quad\text{for } k < m < \omega\,.
\end{align*}
We denote these two logics by $\CMSO$ and $\CGSO$.
\end{Def}

We use a variant of $\MSO_\calP[\Sigma;Q]$ without first-order variables
where the atomic formulae are of the form
\begin{align*}
  X \subseteq Y
  \qtextq{and}
  RZ_0\dots Z_{k-1}\,,
\end{align*}
for relation symbols $R \in \Sigma + \set{ P_q }{ q \in Q }$ and monadic variables
$X,Y,Z_0,\dots,Z_{k-1}$. A~formula of the form~$R\bar Z$ holds if there are elements
$v_i \in Z_i$ such that the tuple~$\bar v$ belongs to the relation~$R$.
It is straightforward to inductively translate every $\MSO$-formula into one of this
special form (see, e.g.,~\cite{Thomas97,BlumensathLN1}).

We will prove below that every $\MSO_\calP$-formula is equivalent to a formula
from a suitable fixed-point logic. For the inductive proof, we will have to
deal with $\MSO_\calP$-formulae~$\varphi(\bar X)$ with free monadic variables~$\bar X$.
In order to make the values of these variables accessible to the fixed-point formula
we annotate each vertex~$v$ of the given tree with the set of variables it belongs to.
Thus, given a tree $t \in \bbT_\bbS Q$ and values $P_0,\dots,P_{n-1} \subseteq \dom(t)$
for~$\bar X$, we construct the tree $t_{\bar P} \in \bbT_\bbS(\PSet(Q) \times \PSet(\bar X))$
with $\dom(t_{\bar P}) = \dom(t)$ and labelling
\begin{align*}
  t_{\bar P}(v) := \langle t(v), U_v\rangle
  \qtextq{where}
  U_v := \set{ X_i }{ v \in P_i }\,.
\end{align*}
The base case of the induction is given by the following lemma.
\begin{Lem}\label{Lem: translation of atomic formulae}
For every atomic\/ $\MSO[\Sigma;Q]$-formula~$\varphi(\bar X)$, there exists
an $\muaf\FO[Q \times \PSet(\bar X)]$-formula~$\psi$ such that
\begin{align*}
  t \models \varphi(\bar P)
  \quad\iff\quad
  t|_{\bar P} \models \psi\,,
  \quad\text{for all $t$ and } \bar P\,.
\end{align*}
\end{Lem}
\begin{proof}
For the labelling of the tree we use the predicates $P_q$~and~$P_X$,
for $q \in Q$ and monadic variables~$X$.
Furthermore, we use the predicates~$P'_\vartheta$, for $\vartheta \in \muaf\FO$,
to check inside of a modal operator~$\medcircle$ whether the corresponding successor satisfies
the $\muaf\FO$-formula~$\vartheta$.

If $\varphi = (X \subseteq Y)$, we use the formula
\begin{align*}
  \psi := \nu x.\bigl[(P_X \lso P_Y) \land {\medcircle}[\forall d. P'_xd]\bigr]\,.
\end{align*}

If $\varphi = P_qX$, for $q \in Q$, we define
\begin{align*}
  \psi := \mu x.\bigl[{\medcircle}[\exists d.P'_xd] \lor (P_X \land P_q)\bigr]\,.
\end{align*}

If $\varphi = (X \preceq Y)$ where $\preceq$~is the tree order on the vertices, we set
\begin{align*}
  \psi := \mu x.\bigl[{\medcircle}[\exists d.P'_xd] \lor \bigl[P_X \land
    \mu y.({\medcircle}[\exists d.P'_yd] \lor P_Y)\bigr]\bigr]\,.
\end{align*}

Finally, if $\varphi = R\bar Z$ where $R$~is one of the relations from the successor structures,
we define
\begin{align*}
  \psi := \mu x.\bigl[{\medcircle}[\exists z.P'_xz] \lor
    {\medcircle}[\exists\bar d.(R\bar d \land \textstyle\Land_i P'_{P_{Z_i}}d_i)]\bigr]\,.
\end{align*}
\upqed
\end{proof}

\begin{Lem}\label{Lem: MSO automata MSO-definable}
Let $\calA = \langle Q,\Sigma,\delta,q_0,\Omega\rangle$ be an automaton.
\begin{enuma}
\item If $\calA$ is a weak $\MSO$-automaton, the language $\lsem\calA\rsem$ is
  $\MSO_{\mathrm{wf}}$-definable.
\item If $\calA$ is a pure $\WMSO_\rmc$-automaton, the language $\lsem\calA\rsem$ is
  $\MSO_{\mathrm{fb}}$-definable.
\item If $\calA$ is a weak $\WMSO_\rmc$-automaton, the language $\lsem\calA\rsem$ is
  $\WMSO$-definable.
\item If $\calA$ is a pure $\FO_\rmd$-automaton, the language $\lsem\calA\rsem$ is
  $\CL$-definable.
\item If $\calA$ is a weak $\FO_\rmd$-automaton, the language $\lsem\calA\rsem$ is
  $\WCL$-definable.
\end{enuma}
\end{Lem}
\begin{proof}
For each state~$q$ of~$\calA$, we will construct a formula~$\varphi_q$ of the
respective logic stating that the given tree has an accepting run which starts
in the state~$q$. We proceed by induction on the number of states reachable from~$q$.
Let $C$~be the connected component of~$\calA$ containing~$q$ and let
$D$~be the set of all states reachable from~$q$ that do not belong to~$C$.
We distinguish two cases.

First suppose that $\delta(q,a) \in L[C,\emptyset]$, where $L$~is the transition logic
in question. We claim that, if there exists an accepting run~$\rho$ with initial state~$q$,
we can choose~$\rho$ such that the set
\begin{align*}
  P := \set{ v }{ \rho(v) \cap C \times C \neq \emptyset }
\end{align*}
forms (a)~a well-founded tree, (b)~a finitely branching tree, (c)~a finite tree, (d)~a chain, or
(e)~a finite chain, respectively.
Then we obtain the desired formula~$\varphi_q$ of the respective logic by stating that
\begin{itemize}
\item there exists a set~$P$ and a family $(Z_{p',p})_{p,p' \in C}$ of sets
  (of the kind allowed by our logic) encoding the restriction of~$\rho$ to~$P$,
\item the root satisfies $Z_{q,q}$,
\item for every infinite branch~$\beta$ that is contained entirely in~$P$,
  every trace along~$\beta$ satisfies the parity condition,
\item the formula
  \begin{align*}
    (\forall v \in P)\Land_{p,p' \in C} [Z_{p',p}v \lso \hat\delta_p(v)]
  \end{align*}
  holds where $\hat\delta_p(v)$~is the formula obtained from~$\delta(p,t(v))$
  by \textsc{(i)}~relativising the formula to the set of successors of~$v$\?;
  \textsc{(ii)}~replacing all states $r \in C$ by the formula $Z_{p,r}$\?;
  and \textsc{(iii)}~replacing all states $r \in D$ by the formula~$\varphi_r$
  (which exists by inductive hypothesis).
\end{itemize}

It therefore remains to prove the above claim.
Fix an accepting run~$\rho$ of~$\calA$ on the given input tree.
We construct an accepting run~$\rho_0$ that has the above property.
We choose $\rho_0(v) \subseteq \rho(v)$ by induction on the distance of~$v$ from the root.

We call an entry $\langle p',p\rangle \in \rho(v)$ \emph{unreachable} if
either $v$~is the root and $p \neq q$, or if $v$~has a predecessor~$u$
and there is no $\langle r',r\rangle \in \rho(u)$ with $r = p'$.
Clearly, (recursively) removing all unreachable entries from an accepting run results
again in an accepting run.
Furthermore note that, if $\rho$~is a run without unreachable entries, the corresponding
set~$P$ is prefix-closed.

(a) If $\calA$~is weak, $\delta(q,a) \in L[C,\emptyset]$ implies that $\Omega(q)$ is odd.
Let $\rho_0$~be the run obtained from~$\rho$ by recursively removing all unreachable
entries. Then the associated set~$P$ does not contain any infinite branches.
Hence, it forms a well-formed tree.

(b) Suppose that $\calA$~is a pure $\MSO_\rmc$-automaton.
For every vertex~$v$ and every entry $\langle p',p\rangle \in \rho(v)$,
it follows that there exists a finite set~$U_{v,p}$ of successors such that
the truth value of the transition formula $\delta(p,t(v))$ does not change
when we remove all entries of the form $\langle p,r\rangle$ from $\rho(w)$,
for successors~$w$ of~$v$ that do not belong to~$U_{v,p}$.
Let $\rho_0$~be the resulting run and $P$~the corresponding set.
It follows that every vertex $v \in P$ has only finitely many successors in~$P$
(those in $\bigcup_{p \in C} U_{v,p}$).
Hence, $P$~forms a finitely-branching tree.

(c) Combining the arguments in (a)~and~(b), we obtain a finite set~$P$.

(d) Given an accpeting run~$\rho$, we will construct an accpeting run~$\rho_0$
and a path $(v_i)_i$ such that each set $\rho_0(v_i)$ contains a single
entry from $C \times C$ while $\rho_0(u)$~is disjoint from $C \times C$,
for all vertices~$u$ that do not lie on the path.

We start with the root~$v_0$ and $\rho_0(v_0) := \{\langle q,q\rangle\}$.
For the inductive step, suppose that we have already defined
$v_i$~and~$\rho_0(v_i)$. Let $\langle p',p\rangle \in \rho_0(v_i)$ be the
unique entry from $C \times C$.
Since the transition formula $\delta(p,t(v_i))$ belongs to $\MSO_\rmd[C,\emptyset]$,
we can find some successor~$u$ of~$v_i$ and some state $r \in C$ such that
the truth value of $\delta(p,t(v_i))$ does not change when we remove all the states from~$C$
from every successer, except for the state~$r$ at~$u$.
If $\langle p,r\rangle \in \rho(u)$, we set
\begin{align*}
  v_{i+1} := u
  \qtextq{and}
  \rho_0(v_{i+1}) :=
    \bigl(\rho(v_{i+1}) \setminus (C \times C)\bigr) \cup \{\langle p,r\rangle\}\,.
\end{align*}
Otherwise, we stop the construction of the path at the vertex~$v_i$.
Finally, we set
\begin{align*}
  \rho_0(u) := \rho(u) \setminus (C \times C)\,,
  \quad\text{for all vertices~$u$ not on the path.}
\end{align*}

The run obtained from~$\rho_0$ by removing all unreachable entries has
the desired properties.

(e) Combining the arguments in (a)~and~(d), we obtain a finite chain~$P$.

\smallskip
It remains to consider the case where $\delta(q,a) \in L[\emptyset,C]$.
Let $\calA^\op$~be the automaton for the complement. Recall that
$\calA^\op$~has the same states as~$\calA$, but their priorities are increased by~$1$,
and the transition formulae are the duals of the formulae from~$\calA$.
Furthermore, a tree is accepted by~$\calA^\op$ with initial state~$q$
if, and only if, the tree is rejected by~$\calA$ with initial state~$q$.
In particular, $C$~is still a component of~$\calA^\op$ and we have
$\delta^\op(q,a) \in L[C,\emptyset]$ for $q \in C$.
Consequently, we can use the above case to find a formula~$\varphi_q$ that defines
the set of all trees that are rejected by~$\calA$ when starting in state~$q$.
The negation~$\neg\varphi_q$ is the desired formula.
\end{proof}

The first characterisation is basically due to Walukiewicz~\cite{Walukiewicz02}.
\begin{Thm}[Walukiewicz~\cite{Walukiewicz02}]\label{Thm: characterisation of MSO}
For a language $K \subseteq \bbT_\bbS\Sigma$, the following statements are equivalent.
\begin{enum1}
\item $K$~is $\MSO$-definable.
\item $K$~is $\mup\MSO$-definable.
\item $K$~is recognised by a pure $\MSO$-automaton.
\end{enum1}
For ordinary trees, the following statements are equivalent to those above.
\begin{enum1}[start=4]
\item $K$~is $\mup\rmE_\omega$-definable.
\item $K$~is recognised by a pure $\rmE_\omega$-automaton.
\end{enum1}
Furthermore, all translations between the above formalisms are effective.
\end{Thm}
\begin{proof}
(3)~$\Leftrightarrow$~(2) follows by
Theorem~\ref{Thm: equivalent automata and fixed-point formulae}.

(2)~$\Rightarrow$~(1) It is straightforward to inductively translate every $\mu\MSO$-formula
into~$\MSO$.

(3)~$\Rightarrow$~(5) One can use a standard back-and-forth
argument~\cite{EbbinghausFlum95,BlumensathLN1} to prove that,
over structures with an empty signature, every $\MSO$-formula
is equivalent to an $\rmE_\omega$-formula.

(5)~$\Rightarrow$~(3) is trivial and
(4)~$\Leftrightarrow$~(5) follows by
Theorem~\ref{Thm: equivalent automata and fixed-point formulae}.

(1)~$\Rightarrow$~(3) By induction on a given $\MSO$-formula~$\varphi$,
we construct an equivalent automaton~$\calA$.
If $\varphi$~is atomic, the existence of~$\calA$ follows by
Lemma~\ref{Lem: translation of atomic formulae}
and the already established implication (2)~$\Rightarrow$~(3).
If $\varphi$~is a boolean combination of $\MSO$-formulae,
the claim follows by inductive hypothesis, the equivalence (2)~$\Leftrightarrow$~(3),
and the fact that the logic $\mup\MSO$ is closed under boolean combinations.
Finally, if $\varphi = \exists X\psi$, the claim follows by inductive hypothesis
and Theorem~\ref{Thm: closure under projections}.
\end{proof}

We obtain analogous results for $\CMSO$, $\GSO$, and $\CGSO$.
The proofs are the same as that of Theorem~\ref{Thm: characterisation of MSO}, except that
for counting logics, we also have to construct automata for predicates of the form
\begin{align*}
  \abs{X} < \infty \,\land\, \abs{X} \equiv k \pmod m\,.
\end{align*}
in the implication (1)~$\Rightarrow$~(3).
\begin{Thm}
For a language $K \subseteq \bbT_\bbS\Sigma$, the following statements are equivalent.
\begin{enum1}
\item $K$~is $\CMSO$-definable.
\item $K$~is $\mup\CMSO$-definable.
\item $K$~is recognised by a pure $\CMSO$-automaton.
\end{enum1}
For ordinary trees, the following statements are equivalent to those above.
\begin{enum1}[start=4]
\item $K$~is $\mup\rmC$-definable.
\item $K$~is recognised by a pure $\rmC$-automaton.
\end{enum1}
Furthermore, all translations between the above formalisms are effective.
\end{Thm}
\begin{Thm}
For a language $K \subseteq \bbT_\bbS\Sigma$, the following statements are equivalent.
\begin{enum1}
\item $K$~is $\GSO$-definable.
\item $K$~is $\mup\GSO$-definable.
\item $K$~is recognised by a pure $\GSO$-automaton.
\end{enum1}
For ordinary trees, the following statements are equivalent to those above.
\begin{enum1}[start=4]
\item $K$~is $\mup\rmE_\omega$-definable.
\item $K$~is recognised by a pure $\rmE_\omega$-automaton.
\item $K$~is $\MSO$-definable.
\end{enum1}
Furthermore, all translations between the above formalisms are effective.
\end{Thm}
\begin{Thm}
For a language $K \subseteq \bbT_\bbS\Sigma$, the following statements are equivalent.
\begin{enum1}
\item $K$~is $\CGSO$-definable.
\item $K$~is $\mup\CGSO$-definable.
\item $K$~is recognised by a pure $\CGSO$-automaton.
\end{enum1}
For ordinary trees, the following statements are equivalent to those above.
\begin{enum1}[start=4]
\item $K$~is $\mup\rmC$-definable.
\item $K$~is recognised by a pure $\rmC$-automaton.
\item $K$~is $\CMSO$-definable.
\end{enum1}
Furthermore, all translations between the above formalisms are effective.
\end{Thm}

For the special case of ordinary trees, the next two theorems are due to~\cite{Carreiro15}.
\begin{Thm}\label{Thm: characterisation of WMSO}
For a language $K \subseteq \bbT_\bbS\Sigma$, the following statements are equivalent.
\begin{enum1}
\item $K$~is $\WMSO$-definable.
\item $K$~is $\muaf\WMSO_\rmc$-definable.
\item $K$~is recognised by a weak $\WMSO_\rmc$-automaton.
\end{enum1}
For ordinary trees, the following statements are equivalent to those above.
\begin{enum1}[start=4]
\item $K$~is $\muaf(\rmE_\infty)_\rmc$-definable.
\item $K$~is recognised by a weak $(\rmE_\infty)_\rmc$-automaton.
\end{enum1}
Furthermore, all translations between the above formalisms are effective.
\end{Thm}
\begin{proof}
The proof is analogous to that of Theorem~\ref{Thm: characterisation of MSO}, except
for the implication (2)~$\Rightarrow$~(1).
Instead, we prove the implication (3)~$\Rightarrow$~(1), which follows
by Lemma~\ref{Lem: MSO automata MSO-definable}\,(c).
\end{proof}

\begin{Thm}\label{Thm: characterisation of WCL}
For a language $K \subseteq \bbT_\bbS\Sigma$, the following statements are equivalent.
\begin{enum1}
\item $K$~is $\WCL$-definable.
\item $K$~is $\muaf\FO_\rmd$-definable.
\item $K$~is recognised by a weak $\FO_\rmd$-automaton.
\end{enum1}
For ordinary trees, the following statements are equivalent to those above.
\begin{enum1}[start=4]
\item $K$~is $\muaf(\rmE_\omega)_\rmd$-definable.
\item $K$~is recognised by a weak $(\rmE_\omega)_\rmd$-automaton.
\end{enum1}
Furthermore, all translations between the above formalisms are effective.
\end{Thm}
\begin{proof}
Again the proof is analogous to that of Theorem~\ref{Thm: characterisation of MSO}, except
for the implication (2)~$\Rightarrow$~(1).
Instead we can prove the implication (3)~$\Rightarrow$~(1) using
Lemma~\ref{Lem: MSO automata MSO-definable}\,(e).
\end{proof}

The following results seem to be new.
\begin{Thm}\label{Thm: characterisation of CL}
For a language $K \subseteq \bbT_\bbS\Sigma$, the following statements are equivalent.
\begin{enum1}
\item $K$~is $\CL$-definable.
\item $K$~is $\mup\FO_\rmd$-definable.
\item $K$~is recognised by a pure $\FO_\rmd$-automaton.
\end{enum1}
For ordinary trees, the following statements are equivalent to those above.
\begin{enum1}[start=4]
\item $K$~is $\mup(\rmE_\omega)_\rmd$-definable.
\item $K$~is recognised by a pure $(\rmE_\omega)_\rmd$-automaton.
\end{enum1}
Furthermore, all translations between the above formalisms are effective.
\end{Thm}
\begin{proof}
Again, we replace the implication (2)~$\Rightarrow$~(1) by (3)~$\Rightarrow$~(1),
which follows by Lemma~\ref{Lem: MSO automata MSO-definable}\,(d).
\end{proof}

\begin{Thm}
For a language $K \subseteq \bbT_\bbS\Sigma$, the following statements are equivalent.
\begin{enum1}
\item $K$~is $\MSO_{\mathrm{fb}}$-definable.
\item $K$~is $\mup\WMSO_\rmc$-definable.
\item $K$~is recognised by a pure $\WMSO_\rmc$-automaton.
\end{enum1}
For ordinary trees, the following statements are equivalent to those above.
\begin{enum1}[start=4]
\item $K$~is $\mup(\rmE_\infty)_\rmc$-definable.
\item $K$~is recognised by a pure $(\rmE_\infty)_\rmc$-automaton.
\end{enum1}
Furthermore, all translations between the above formalisms are effective.
\end{Thm}
\begin{proof}
Again the proof is similar to the ones above.
For (3)~$\Rightarrow$~(1), we can use Lemma~\ref{Lem: MSO automata MSO-definable}\,(b).
\end{proof}

\begin{Thm}
For a language $K \subseteq \bbT_\bbS\Sigma$, the following statements are equivalent.
\begin{enum1}
\item $K$~is $\MSO_{\mathrm{wf}}$-definable.
\item $K$~is $\muaf\MSO$-definable.
\item $K$~is recognised by a weak $\MSO$-automaton.
\end{enum1}
For ordinary trees, the following statements are equivalent to those above.
\begin{enum1}[start=4]
\item $K$~is $\muaf\rmE_\omega$-definable.
\item $K$~is recognised by a weak $\rmE_\omega$-automaton.
\end{enum1}
Furthermore, all translations between the above formalisms are effective.
\end{Thm}
\begin{proof}
Again the proof is similar to the ones above.
For (3)~$\Rightarrow$~(1), we can use Lemma~\ref{Lem: MSO automata MSO-definable}\,(1).
\end{proof}

\begin{Open}
Is there an automaton characterisation of $\MSO^\calP$ where $\calP$~is the set of thin trees\??
\end{Open}

\section{The Muchnik Iteration}   
\label{Sect: Muchnik}

As an application of the machinery developed in this article we consider the so-called
Muchnik iteration which, given some $\Sigma$-structure~$\frakA$, creates an infinite tree
of copies of~$\frakA$. It can be seen as a generalisation of the unravelling operation
to arbitrary $\Sigma$-structures.
\begin{Def}
Let $\frakA = \langle A,\bar R\rangle$ be a $\Sigma$-structure.

(a)
The \emph{Muchnik iteration} of~$\frakA$ is the
$(\Sigma + \{E,\cl\})$-structure $\frakA^* = \langle A^*,E,\cl,\bar R^*\rangle$
whose universe consists of all finite sequence over~$A$ and
\begin{align*}
  R_i^* &:= \set{ \langle wa_0,\dots,wa_{n-1}\rangle }{ w \in A^*,\ \bar a \in R }\,, \\
  E     &:= \set{ \langle w,wa\rangle }{ w \in A^*,\ a \in A }\,, \\
  \cl   &:= \set{ waa }{ w \in A^*,\ a \in A }\,.
\end{align*}
We call $\cl$ the \emph{clone predicate.}

(b) We regard~$\frakA^*$ as an $\bbS$-enriched transition system for the functor $\bbS X := X^A$
by choosing for $\suc(w)$ (with $w \in A^*$) the substructure of~$\frakA^*$ induced by the set
\begin{align*}
  \set{ wa }{ a \in A }\,.
\end{align*}
\upqed
\end{Def}

The following theorem, originally due to Muchnik, is one of the strongest decidability results
for $\MSO$ known. The theorem was announced in~\cite{Semenov84}, but the original proof
has never been published. Walukiewicz~\cite{Walukiewicz02} introduced $\MSO$-automata
to give a new independent proof of this theorem.
Here we present a new, much simplified proof that also applies to several other logics.
\begin{Thm}[Muchnik, Walukiewicz~\cite{Walukiewicz02}]\label{Thm: Muchnik}
Given an\/ $\MSO$-formula~$\varphi$, we can compute an\/ $\MSO$-formula~$\varphi^*$ such that
\begin{align*}
  \frakA^* \models \varphi \quad\iff\quad \frakA \models \varphi^*\,,
  \quad\text{for all structures } \frakA\,.
\end{align*}
\end{Thm}

The proof of this theorem is split into two lemmas. Let us start with a bit of terminology.
\begin{Def}
(a)
A \emph{system} of logics is a functor~$L$ mapping each finite relational signature~$\Sigma$
to a logic $L[\Sigma]$ whose class of models is the class of all $\Sigma$-structures.

(b)
Given such a system~$L$ and a signature~$\Sigma$, we construct a family of logics~$L_\Sigma$ by
\begin{align*}
  L_\Sigma[Q] := L[\Sigma + \set{ P_q }{ q \in Q }]\,,
  \quad\text{for every set} Q\,,
\end{align*}
where the~$P_q$ are unary predicates.
\end{Def}
\begin{Rem}
(a)
I~apologise for any confusion caused by defining both families of logics and systems of logics,
but I~was simply not able to come up with better terminology.

(b) Clearly, all of the classical logics like first-order logic, monadic second-order logic,
etc.\ are systems of logics.
\end{Rem}

Next, let us extend the logic~$\mu L$ by a mechanism for \emph{loop-detection.}
\begin{Def}
Let $L$~be a family of logics with polarities over $\bbS \circ \PSet$.

(a) We denote by $\mu^\rcirclearrowleft L$ the following variant of~$\mu L$.
Given a set~$\Sigma$ of labels and two disjoint sets $X,Y$ of fixed-point variables,
we define the syntax and semantics of $\mu^\rcirclearrowleft L[\Sigma;X,Y]$ using the same
rules for as $\mu L[\Sigma;X,Y]$, except the one for the modal operators, which is modified
as follows. In the syntax, we allow an additional label ${*} \in \one$.
\begin{itemize}
\item Let $\Theta \subseteq \mu L[\Sigma;X,Y]$ be a finite set of formulae
  and let $\Theta_+$~be the set of all $\vartheta \in \Theta$ containing a symbol from~$X$ and
  $\Theta_-$~the set of all $\vartheta \in \Theta$ containing a symbol from~$Y$.
  For every $\varphi \in L[\Theta+\one,\Theta_+,\Theta_-]$, we have
  ${\medcircle}\varphi \in \mu L[\Sigma;X,Y]$.
\end{itemize}
The label~$*$ is used to mark loops, that is, if the vertex~$v$ is a successor of itself,
we label it by~$*$. Formally, we set
\begin{align*}
  \lsem{\medcircle}\psi\rsem_{\bar P} :=
    \bigset{ v \in S }{ \bbS f_v(\suc(v)) \models \psi }
\end{align*}
where the function $f_v : S \to \PSet(\Theta)$ maps $u \in S$ to
\begin{align*}
  \set{ \vartheta \in \Theta }{ u \in \lsem\vartheta\rsem_{\bar P} }
  \cup
  \begin{cases}
    \one      &\text{if } u = v\,, \\
    \emptyset &\text{otherwise}\,,
  \end{cases}
\end{align*}

(b) We denote by $\mup^\rcirclearrowleft L$ and $\muaf^\rcirclearrowleft L$ the
corresponding \emph{pure} and \emph{alternation-free} fragments of $\mu^\rcirclearrowleft L$.
\end{Def}

\begin{Def}
Given a $\Sigma$-structure~$\frakA$, we denote by~$\widehat\frakA$ the
$(\Sigma + \{E\})$-structure $\langle\frakA + \one,E\rangle$ with edge
relation $E := (A + \one) \times A$.
\end{Def}
Note that the unravelling of~$\widehat\frakA$ coincides with~$\frakA^*$,
except that it does not have the clone predicate.

The first step in the proof of Theorem~\ref{Thm: Muchnik} consists in proving
a variant for the logic~$\mu^\rcirclearrowleft L$.
\begin{Lem}\label{Lem: from A* to hat A}
Let $L$~be a system of logics with polarities over $\bbS \circ \PSet$.
For every $\mu L$-formula~$\varphi$, there exists a $\mu^\rcirclearrowleft L$-formula~$\varphi^*$
such that
\begin{align*}
  \frakA^* \models \varphi \quad\iff\quad \widehat\frakA \models \varphi^*,
  \quad\text{for every $\Sigma$-structure } \frakA\,.
\end{align*}
Furthermore, if $\varphi \in \mup L$ or $\varphi \in \muaf L$,
we can choose $\psi \in \mup^\rcirclearrowleft L$ and $\psi \in \muaf^\rcirclearrowleft L$,
respectively.
\end{Lem}
\begin{proof}
Let $\varphi(\bar x)$ be a $\mu L$-formula, possibly with free fixed-point variables~$\bar x$.
By induction on~$\varphi$, we construct an $\mu^\rcirclearrowleft L$-formula~$\varphi^*(\bar x)$
such that
\begin{align*}
  \frakA^* \models \varphi(\rho^{-1}[\bar P])
  \quad\iff\quad
  \widehat\frakA \models \varphi^*(\bar P)\,,
\end{align*}
for every $\Sigma$-structure~$\frakA$ and all sets~$\bar P$ in~$\widehat A$,
where $\rho : A^* \to \widehat A = A + \one$ is the function mapping the empty word~$\emptyseq$
to the unique element ${*} \in \one$ and every other word to its last letter.
The only two non-trivial steps in the induction is the case of modal operators and fixed-points.

First, suppose that $\varphi = {\medcircle}\psi$.
Fix a finite set~$\Theta$ of $\mu L$-formulae such that
$\psi \in  L_{\Sigma + \{\cl\}}[\Theta]$.
By inductive hypothesis, we can translate every $\vartheta \in \Theta$ into an equivalent
$\mu^\rcirclearrowleft L$-formula~$\vartheta^*$.
Let $\Theta^*$~be the resulting set and let $\psi^*$~be the formula obtained from~$\psi$
by replacing each~$\vartheta$ by~$\vartheta^*$.
Since
\begin{align*}
  L_{\Sigma + \{\cl\}}[\Theta] = L[\Sigma + \{\cl\} + \Theta] = L_\Sigma[\Theta + \{\cl\}],
\end{align*}
it follows that $\varphi^* := {\medcircle}\psi^* \in \mu^\rcirclearrowleft L$.
Furthermore, the fact that
\begin{align*}
  \frakA^* \models \varphi(\rho^{-1}[\bar P])
  \quad\iff\quad
  \widehat\frakA \models \varphi^*(\bar P)
\end{align*}
follows immediately from the inductive hypothesis.

It remains to consider a fixed-point formula $\mu y.\psi(\bar x,y)$.
Let $\psi^*$~be the $\mu^\rcirclearrowleft L$-formula obtained by inductive hypothesis.
We set $\varphi^* := \mu y.\psi^*$. To see that $\varphi^*$~has the desired properties,
let $(Q_i)_i$ be the fixed-point induction of~$\psi$ on the structure~$\frakA^*$.
By induction on~$i$, it then follows that $Q_i = \rho^{-1}[P_i]$, where $P_i$~is the $i$-th
stage of the fixed-point induction of~$\psi^*$ on~$\widehat\frakA$.
\end{proof}

Finally, we have to translate $\mu^\rcirclearrowleft L$ back into the logics we are actually
interested in.
\begin{Lem}\label{Lem: mu L to L}\leavevmode
\begin{enuma}
\item Let $L$~be\/ $\MSO$ or\/ $\GSO$.
  For every $\mu^\rcirclearrowleft L$-formula~$\varphi$, there exists an $L$-formula~$\psi$
  such that
  \begin{align*}
    \frakS \models \varphi \quad\iff\quad \frakS \models \psi\,,
    \quad\text{for every transition system } \frakS\,.
  \end{align*}
\item For every $\muaf^\rcirclearrowleft \WMSO_\rmc$-formula~$\varphi$, there exists
  a\/ $\WMSO$-formula~$\psi$ such that
  \begin{align*}
    \frakS \models \varphi \quad\iff\quad \frakS \models \psi\,,
    \quad\text{for every transition system } \frakS\,.
  \end{align*}
\end{enuma}
\end{Lem}
\begin{proof}
(a)
Given a $\mu^\rcirclearrowleft L$-formula~$\varphi(\bar x)$, possibly with free fixed-point
variables~$\bar x$, we inductively construct an $L$-formula~$\psi(z,\bar X)$ such that
\begin{align*}
  \frakS,s \models \varphi(\bar P) \quad\iff\quad \frakS \models \psi(s,\bar P)\,,
\end{align*}
for all transition systems~$\frakS$ and all sets~$\bar P$.
Most steps of the induction are trivial. For a fixed point $\mu y.\varphi(\bar x,y)$
we can express in~$L$ that $z \in Y$, where $Y$~is the least set satisfying
\begin{align*}
  \forall y[y \in Y \liff \psi(y,\bar X,Y)]\,.
\end{align*}
For a modal operator~${\medcircle}\vartheta$, we can use the relativisation of the
formula~$\vartheta$ to the set~$U$ of all successors of~$z$. The additional
predicate used for loop-detection is equal to $U \cap \{z\}$, which is obviously definable.

(b) We proceed analogously to~(a), the only exception being the translation of
fixed-point operators.
Hence, suppose that $\varphi = \mu x_0\cdots\mu x_{n-1}\psi(\bar x)$
where $\psi \in \WMSO_\rmc[\bar x,\bar x,\emptyset]$ is continuous in~$\bar x$.
By induction on~$n$ we prove that, for every transition system~$\frakS$,
there exists a finite set $U \subseteq S$ such that
\begin{align*}
  \frakS \models \mu x_0\cdots\mu x_{n-1}\psi
  \quad\iff\quad
  \frakS \models \mu x_0\cdots\mu x_{n-1}[U \land \psi]\,.
\end{align*}
Then it follows that we can define the fixed-point in~$\WMSO$ by stating that
$z \in Y$, where $Y \subseteq U$ is the least set satisfying
\begin{align*}
  \forall y[y \in Y \liff y \in U \land \psi(y,\bar X,Y)]\,.
\end{align*}

First, suppose that $n = 1$. Let $F : \PSet(S) \to \PSet(S)$ be the function associated with
the formula~$\psi$, let $H_i := F^i(\emptyset)$ be the $i$-th stage of the fixed-point
induction, and let $H_\infty$~be the limit.
Since $\psi$~is continuous in~$x_0$, it follows that, for every $i < \omega$ and every
element $u \in H_{i+1}$, there exists some finite set $W_u \subseteq H_i$ with $u \in F(W_u)$.

Let us show that this implies that $H_\infty = H_\omega$.
For a contradiction, suppose otherwise. Then there exists some element
$u \in H_{\omega+1} \setminus H_\omega$.
Let $k$~be the maximal number such that $W_u$~contains some element of $H_{k+1} \setminus H_k$.
Since $W_u$~is finite, it follows that $k < \omega$.
Hence, $u \in F(W_u) \subseteq F(H_k) \subseteq H_\omega$. A~contradiction.

Setting
\begin{align*}
  U_u := \{u\} \cup \bigcup_{w \in W_u} U_w\,,
\end{align*}
it now follows by induction on~$i$ that,
for every $u \in H_i$, there is some finite set $U_u \subseteq S$ with
\begin{align*}
  \frakS,u \models \mu x_0[U_u \land \psi]\,.
\end{align*}
Hence, the set $U := U_v$ has the desired properties.

For the inductive step, suppose that $n > 1$.
We can use the above case to find a finite set~$W$ such that
\begin{align*}
  \frakS \models \mu x_0[W \land \mu x_1\cdots\mu x_{n-1}\psi]\,.
\end{align*}
Furthermore, we can use the inductive hypothesis to find, for every $w \in W$
and every finite $P \subseteq W$, some finite set~$V_{P,w}$ with
\begin{align*}
  &\frakS,w \models \mu x_1\cdots\mu x_{n-1}\psi(P,x_1,\dots,x_{n-1}) \\
  \iff\quad
  &\frakS,w \models \mu x_1\cdots\mu x_{n-1}[V_{P,w} \land \psi(P,x_1,\dots,x_{n-1})]\,.
\end{align*}
Setting $U := \bigcup_{P,w} V_{P,w}$ it follows that
\begin{align*}
  \frakS,u \models \mu x_1\cdots\mu x_{n-1}[U \land \psi(H_i,x_0,\dots,x_{n-1})]\,,
  \quad\text{for all } u \in W \cap H_{i+1}\,.
\end{align*}
Consequently, we have
\begin{align*}
  \frakS \models \mu x_0\cdots\mu x_{n-1}[U \land \psi]\,.
\end{align*}
\upqed
\end{proof}

Combining these two lemmas we obtain the following theorem.
The case $L = \MSO$ is the origninal Theorem of Muchnik,
the cases for $\CMSO$, $\GSO$, and $\CGSO$ were proved in~\cite{BlumensathKreutzer05},
and the case $L = \WMSO$ is new.
\begin{Thm}
Let $L$~be one of\/ $\MSO$, $\CMSO$, $\GSO$, $\CGSO$, or\/~$\WMSO$.
Given an\/ $L$-formula~$\varphi$, we can compute an\/ $L$-formula~$\varphi^*$ such that
\begin{align*}
  \frakA^* \models \varphi \quad\iff\quad \frakA \models \varphi^*\,,
  \quad\text{for all structures } \frakA\,.
\end{align*}
\end{Thm}
\begin{proof}
We give the proof for $L = \MSO$. The other cases are entirely analogous.
We can use Theorem~\ref{Thm: characterisation of MSO} to translate a given
$\MSO$-formula~$\varphi$ into a $\mu\MSO$-formula~$\psi$.
Let $\psi^*$~be the $\mu^\rcirclearrowleft\MSO$-formula obtained from~$\psi$
via Lemma~\ref{Lem: from A* to hat A}.
We obtain the desired $\MSO$-formula~$\varphi^*$ by applying Lemma~\ref{Lem: mu L to L}
to~$\psi^*$.
\end{proof}

{\small\raggedright
\bibliographystyle{siam}
\bibliography{Automata-submitted.bib}}

\end{document}